\documentclass[journal=jctcce,manuscript=article,layout=two-column]{achemso}

\usepackage[T1]{fontenc}
\usepackage[utf8]{inputenc}
\usepackage{amsmath,amssymb,bm}
\usepackage{booktabs}
\usepackage{array}
\usepackage{multirow}
\usepackage{tabularx}
\usepackage{longtable}
\usepackage{siunitx}
\usepackage{xcolor}
\usepackage{enumitem}
\usepackage{flushend}
\usepackage[colorlinks=true,linkcolor=black,citecolor=black,urlcolor=blue]{hyperref}

\setlist{nosep,leftmargin=1.8em,topsep=0.25em}
\newcolumntype{L}[1]{>{\raggedright\arraybackslash}p{#1}}
\newcommand{\tabfont}{\footnotesize}
\newcommand{\code}[1]{\texttt{#1}}
\newcommand{\doi}[1]{\href{https://doi.org/#1}{\texttt{doi:#1}}}

\title[Affordable analytic $G_0W_0$-BSE gradients]
{Affordable Analytic Gradients for $G_0W_0$-BSE Excited States for Multiple $GW$ Variants via Back-Propagation}

\author{Qirui Gao}
\affiliation{Research Center for Chemical Theory, State Key Laboratory of Porous Materials for Separation and Conversion, Collaborative Innovation Center of Chemistry for Energy Materials, Shanghai Key Laboratory of Molecular Catalysis and Innovative Materials, MOE Key Laboratory of Computational Physical Sciences, Department of Chemistry, Fudan University, Shanghai 200433, China}
\author{Igor Ying Zhang}
\email{igor\_zhangying@fudan.edu.cn}
\affiliation{Research Center for Chemical Theory, State Key Laboratory of Porous Materials for Separation and Conversion, Collaborative Innovation Center of Chemistry for Energy Materials, Shanghai Key Laboratory of Molecular Catalysis and Innovative Materials, MOE Key Laboratory of Computational Physical Sciences, Department of Chemistry, Fudan University, Shanghai 200433, China}
\affiliation{Hefei National Laboratory, Hefei 230088, China}
\author{Xin Xu}
\affiliation{Research Center for Chemical Theory, State Key Laboratory of Porous Materials for Separation and Conversion, Collaborative Innovation Center of Chemistry for Energy Materials, Shanghai Key Laboratory of Molecular Catalysis and Innovative Materials, MOE Key Laboratory of Computational Physical Sciences, Department of Chemistry, Fudan University, Shanghai 200433, China}
\affiliation{Hefei National Laboratory, Hefei 230088, China}


\hypersetup{
pdftitle={Affordable analytic gradients for G0W0-BSE excited states for multiple GW variants via back-propagation},
pdfauthor={Qirui Gao, Igor Ying Zhang, Xin Xu},
pdfsubject={Journal of Chemical Theory and Computation},
pdfkeywords={GW approximation, Bethe-Salpeter equation, analytic nuclear gradients, excited-state geometry optimization, nonadiabatic molecular dynamics}
}

\makeatletter
\renewcommand\normalsize{\@setfontsize\normalsize{10}{11}}
\renewcommand\small{\@setfontsize\small{9}{10}}
\renewcommand\footnotesize{\@setfontsize\footnotesize{8}{9}}
\renewcommand\scriptsize{\@setfontsize\scriptsize{7}{8}}
\makeatother
\AtBeginDocument{\setstretch{1.5}\normalsize}

\begin{document}

\begin{abstract}
We derive, implement and validate analytic nuclear gradients for three variants of
$GW$ self-energies---the contour-deformation (CD), the analytic continuation (AC), and the contour deformation with $W$-analytic-continuation (CD-WAC)---and for the
Bethe--Salpeter equation (BSE) built on top of them. All gradients are expressed as
contractions of a set of covectors connected to derivatives of primitive quantities in SCF calculations. The derivation of these covectors is a back-propagation procedure through the $GW$ and BSE equations.
Central finite differences confirm the correctness of all implemented gradients. With the AC-$GW$-BSE gradient we optimize the excited-state geometries of carbon monoxide, a donor-acceptor dye labelled as DA1, and DMABN, a typical model system for twisted intramolecular charge-transfer phenomena (TICT), and have obtained results that successfully reproduce observations from experiments or high-level calculations. We also performed trajectory surface hopping (TSH) simulations of nonadiabatic molecular dynamics (NAMD) of DMABN using the SHARC NAMD package, showing the capacity of our method to simulate NAMD problems. We performed timing and memory scaling tests, demonstrating that the gradient calculations of both $GW$ and BSE scale asymptotically as $N^4$ for timing and $N^3$ for memory.
All source codes used in the present work, including the REST program package and the modified SHARC NAMD package that supports REST as the quantum chemistry backend, are available on Gitee (\url{https://gitee.com/restgroup/rest} and
\url{https://gitee.com/qirui-gao-2025/sharc}).
\end{abstract}

\section{\raggedright\large Introduction}

Potential-energy surfaces of electronically excited states are the working objects of
photochemistry, photobiology and the computational design of photoactive materials.
Optimizing excited-state geometries
requires the gradient of the excited-state energy with respect to the nuclear
coordinates, and the availability of an analytic gradient makes such searches affordable
for polyatomic molecules \cite{pulay1969,schlegel1982,bernardi1996}. Analytic gradients
are also the enabling element of nonadiabatic molecular dynamics (NAMD), where the
gradient is needed at every time step.

For molecules of moderate size the method of choice for excited-state gradients has long been time-dependent density functional theory (TD-DFT) within the adiabatic approximation. Analytic geometric derivatives of TD-DFT excitation energies were
formulated by Van Caillie and Amos \cite{vancaillie1999} and put on a general
response-theory footing by Furche and Ahlrichs \cite{furche2002}, with later extensions
to solvation \cite{scalmani2006} and to nonadiabatic couplings \cite{send2010}. These
implementations have been applied very successfully to problems including the
photoisomerization mechanisms of azobenzene \cite{cembran2004}, the excited-state
dynamics of photochemically active molecules \cite{tapavicza2007}, and a broad range of
spectroscopic and photophysical problems \cite{dreuw2005,furche2005}. However, the
well-known limitations of TD-DFT in charge-transfer and Rydberg excitations still pose
challenges to many NAMD studies where such excitations are the excitations of interest.

The $GW$-BSE approach has been an emerging alternative approach to TD-DFT which is known
for describing charge-transfer and Rydberg excitations with same accuracy as local
excitations. In the $G_0W_0$ approximation the quasiparticle (QP) energies are obtained
from the $GW$ self-energy, and the same self-energy enters the electron--hole interaction kernel when
solving the Bethe--Salpeter equation (BSE) \cite{onida2002,rohlfing2000}. The resulting
$GW$-BSE procedure reproduces optical excitation energies of molecules with an accuracy that
is competitive with high-level wavefunction methods \cite{blase2018,jacquemin2015}, and
it has become a increasingly important tool for the excited states of moleculaes and clusters.

Excited-state geometries within $GW$-BSE have so far been obtained either with analytic
gradients that carry additional approximations, including an excited state force of
$GW$-BSE derived by Ismail-Beigi and Louie \cite{ismail2003}, Lagrangian $Z$-vector BSE
gradients assessed under further approximations derived by Villalobos-Castro
\emph{et al.} \cite{villalobos2023}, and a corrected force scheme that retains an
approximate $GW$-level electron--phonon kernel \cite{delgrande2025}, or with
computational schemes that include elements of numerical gradients from finite
differences, including an excited state force evaluation scheme that utilizes the finite
difference of an excitonic Hamiltonian \cite{alrahamneh2025}, and a study that employs
numerical forces to obtain excited state geometries \cite{caylak2021}. The only complete,
approximation-free formulation is that of T\"olle and co-workers, for $G_0W_0$
\cite{tolleG0W0} and for BSE@$G_0W_0$ \cite{tolle2025}, but its $O(N^6)$ cost and its
incompatibility with the contour-deformation (CD-$GW$) \cite{golze2018} and
analytic-continuation (AC-$GW$) \cite{ren2012} frameworks of mainstream $GW$-BSE codes make it difficult to generalize and
apply. In the absence of an affordable alternative, the excited-state surfaces of
molecules within $GW$-BSE have therefore been mapped mainly by scanning a few nuclear
coordinates, one energy evaluation at a time \cite{faber2014,knysh2022,knysh2023}, which
underlines the need for an efficient analytic gradient inside these frameworks.
During the final revision of this manuscript, we became aware of a very recent
preprint by Holzer \cite{holzer2026} presenting a TD-DFT-like $Z$-vector formulation of
analytic $GW$-BSE gradients, which is also free of additional approximations. Nevertheless, the present work develops a covector-based back-propagation derivation which is substantially different from the $Z$-vector-centered route of Holzer's work,
and it is potentially inspiring for future derivation and implementation of analytic gradients of other quantum chemical methods. We therefore believe that this work will also be of interest to the
theoretical chemistry community.

Traditional analytic gradients in TD-DFT are usually derived by constructing a global
Lagrangian and introduce Lagrange multipliers for the
orbital response\cite{furche2002,scalmani2006}. The resulting gradient is expressed
through relaxed one- and two-particle density matrices and a $Z$-vector obtained from
the coupled-perturbed Kohn--Sham equations(CP-KS); the derivation is thus organized around the
stationarity conditions of a method-specific functional, with the $Z$-vector as the
central auxiliary object.

The present work is organized around a different decomposition. We split every
analytic gradient into two stages: (i) the gradient of the post-SCF target---here the QP
energy or the BSE excitation energy---is back-propagated into a form expressed by the derivatives of a small set of physical quantities that already appear in the standard SCF calculation; (ii) the derivatives
of these SCF quantities with respect to the nuclear coordinates are evaluated once,
independently of the post-SCF method. We call the quantities that are differentiated in stage (ii) \textbf{primitives}.
In the present implementation they are the orbital energies $\epsilon_p$, the RI
three-centre tensor $I_{P,pq}$, the Coulomb metric $J_{PQ}$, and the diagonal
exchange-correlation potential $v^{xc}_p$; the MO coefficients enter through the orbital
rotations that accompany the primitive derivatives. The first part is obtained by
reverse-mode back-propagation: the QP energy or the BSE eigenvalue is treated as a
scalar output of a network of primitives and intermediate quantities, and the chain rule
is applied locally to the correlation self-energy, the screened interaction, the BSE
kernel, and the exchange-correlation potential. This yields \textbf{covectors} on the primitives,
which are contracted with the primitive derivatives from the second stage. While the CP-KS equations are still solved, they act only as the
back-propagation through the orbital-rotation node rather than as the central organizing
principle of the derivation.

One reason the architecture of the present work is particularly interesting lies in the development of automatic differentiation (AD) frameworks for quantum chemistry in recent years, both as fully differentiable program packages \cite{zhang2022,song2021} and as tools for deriving analytic gradients of complicated wavefunction models \cite{mitchell2024}. By utilizing automatically differentiable codes, analytic gradients of various quantities should be in principle straightforward. However, it remains challenging to develop computationally-efficient and scalable back-propagation frameworks for quantum chemical methods. While this work still derives the analytic gradients manually instead of using AD, it uses techniques highly related to AD and shares the same mathematical foundation, namely back-propagation and covector assembly. This work therefore provides a clear example of constructing the analytic gradients using back-propagation techniques while preserving the asymptotic computational scaling of the original quantum chemical method. This may potentially become useful reference for future development of AD frameworks and other quantum chemical methods.

In this work we derive and implement, within the REST program \cite{rest2025,restgwbse2026},
analytic nuclear gradients for three $GW$ variants --- contour-deformation
CD-$GW$\cite{golze2018}, the analytic-continuation AC-$GW$\cite{ren2012}, and the $W$-analytically-continued contour deformation(CD-WAC)\cite{panades2023}---and for the BSE excitation
energies built on top of them. The computational cost of gradients of both $GW$ and BSE scales as $O(N^4)$ and the
memory requirement as $O(N^3)$ with system size, and the formulation can be transferred to
any program that implements CD-$GW$, AC-$GW$ or CD-WAC-$GW$. The gradient is expressed through a small
set of primitive quantities---the orbital energies, the RI three-centre tensor, the
Coulomb metric, and the diagonal exchange-correlation potential---so that the QP part and
the BSE kernel
part share one set of primitive gradients and one folding/adjoint machinery, and the $GW$
variant enters only through a set of frequency weights.

The paper is organized as follows. Sec.~2 presents the theory: a review of $GW$ and BSE(Sec.~2.1), the analytic gradient of the
AC-$GW$ quasiparticle energy including the differentiation of the Pad\'e continuation
(Sec.~2.2), the BSE analytic gradient expressed as a $GW$-independent primitive
contraction (Sec.~2.3) and the primitive gradients (Sec.~2.4). The analytic gradients under the CD and
CD-WAC schemes, whose frameworks are highly similar to Sec.~2.2, are derived in the
Supporting Information. Sec.~3 summarizes the computational setup and Sec.~4 shows the
numerical validation: the finite-difference validation of the three $GW$ variants and of
the AC-$GW$-BSE gradient (Sec.~4.1), the excited-state structures of several molecules
(Sec.~4.2), an example of NAMD simulation using the SHARC package\cite{sharc2018,sharc2026} (Sec.~4.3), and the cost and memory scaling
(Sec.~4.4).

\section{\raggedright\large Theory}

Throughout this section atomic units are used, real orbitals are assumed, and
$i,j$ label occupied, $a,b$ virtual, $m,n,p,q$ generic, and $\mu,\nu$
atomic orbitals. $P,Q$ label auxiliary functions (used in
the RI approximation). $k$ labels the imaginary-frequency quadrature nodes. $\xi$ denotes a nuclear coordinate.

\subsection{\raggedright\normalsize Review of $GW$ and of the Bethe--Salpeter equation}

\paragraph{The $GW$ quasiparticle equation.}
Within the diagonal approximation, the quasiparticle energy of orbital $n$ is the root of
\begin{equation}
\omega_n=\epsilon_n+\Sigma^x_{nn}-v^{xc}_{n}
+\mathrm{Re}\,\Sigma^c_{nn}(\omega_n),
\end{equation}
where $\epsilon_n$ is the orbital energy from the preceding SCF calculation (Hartree--Fock
\cite{hartree1928,fock1930,roothaan1951} or a density functional
\cite{perdew1996,adamo1999}), $\Sigma^x$ the exact-exchange self-energy, and $v^{xc}$ the
exchange--correlation potential of the density functional;
$v^{xc}_p\equiv v_{xc,pp}$ denotes its diagonal matrix element. The
correlation self-energy is expressed through the RPA screened interaction
\cite{golze2018},
\begin{equation}
W^c_{pq}(z)=\sum_{P,Q}I_{P,pq}\big[(J-Q(z))^{-1}-J^{-1}\big]_{PQ}I_{Q,pq},
\end{equation}
built from the RI three-centre integrals and the Coulomb metric
\begin{equation}
\begin{aligned}
I_{P,pq}&=\sum_{\mu\nu}I_{\mu\nu P}C_{\mu p}C_{\nu q},\\
J_{PQ}&=(P|Q),\\
Q_{PQ}(z)&=\sum_{ia}I_{P,ia}\,d_{ia}(z)\,I_{Q,ia},
\end{aligned}
\end{equation}
with the independent-particle response at real or imaginary frequencies,
\begin{equation}
\begin{aligned}
d_{ia}(\text{i}u)&=-\frac{4\Delta_{ia}}{\Delta_{ia}^{2}+u^{2}},\\
d_{ia}(\omega)&=-\frac{4\Delta_{ia}}{\Delta_{ia}^{2}-\omega^{2}},\\
\Delta_{ia}&=\epsilon_a-\epsilon_i.
\end{aligned}
\end{equation}

\paragraph{The self-energy on the imaginary axis.}
In the imaginary-axis (AC) variant \cite{ren2012} the exact self-energy is evaluated only at
imaginary frequencies, where it is a smooth analytic function,
\begin{equation}
\begin{aligned}
\Sigma^{c}_{nn}(\text{i}u)&=-\frac{1}{\pi}\int_0^{\infty}\!\frac{t_n}{t_n^{2}+u^{2}}\,W^c_{nn}(\text{i}u')\,\text{d}u'\Big|_{u'=u},\\
t_n&=\text{i}u-\epsilon_n,
\end{aligned}
\end{equation}
and the self-energy at real $\omega$ is obtained by analytic
continuation of the sampled function,
\begin{equation}
\Sigma^{c}_{nn}(\omega)\simeq\mathcal{P}\big[\{\Sigma^{c}_{nn}(\xi_k)\}\big](\omega),
\end{equation}
where $\mathcal{P}$ is a Pad\'e approximant built from samples taken on the Fermi-shifted
imaginary axis, $\xi_k=e_F+\text{i}\lambda_k$ with
$e_F=\tfrac12(\epsilon_{\rm HOMO}+\epsilon_{\rm LUMO})$, at the fixed quadrature nodes
$\{\lambda_k\}$ \cite{vidberg1977,ren2012}. In practice the integral of
Eq.~(5) is discretized with a Gauss--Legendre quadrature on the semi-infinite interval,
\begin{equation}
\Sigma^{c}_{nn}(\text{i}u)\simeq\sum_k w_k\,\frac{t_n}{t_n^{2}+u_k^{2}}\,W^c{nn}(\text{i}u_k).
\end{equation}

\paragraph{The Bethe--Salpeter equation.}
The BSE excitation energies $\Omega_S$ and amplitudes $(X_S,Y_S)$ solve the Casida-like eigenvalue problem \cite{rohlfing2000,blase2018},
\begin{equation}
\begin{gathered}
\begin{pmatrix}A&B\\ B&A\end{pmatrix}
\begin{pmatrix}X_S\\ Y_S\end{pmatrix}
=\Omega_S
\begin{pmatrix}1&0\\ 0&-1\end{pmatrix}
\begin{pmatrix}X_S\\ Y_S\end{pmatrix},\\
X_S^{T}X_S-Y_S^{T}Y_S=1,
\end{gathered}
\end{equation}
with the blocks
\begin{equation}
\begin{aligned}
A_{ia,jb}&=\delta_{ij}\delta_{ab}(E^{GW}_a-E^{GW}_i)+\kappa\,v_{ia,jb}-W^{A}_{ij,ab},\\
B_{ia,jb}&=\kappa\,v_{ia,jb}-W^{B}_{ib,ja},
\end{aligned}
\end{equation}
in which the $GW$ quasiparticle energies $E^{GW}_p$ dress the diagonal, $\kappa=2$ ($0$) for
the singlet (triplet) channel, and the kernel is built from the same primitive quantities as the $GW$ self-energy and applies the static approximation $z=0$,
\begin{equation}
\begin{aligned}
v_{ia,jb}&=\sum_{P,Q}I_{P,ia}(J^{-1})_{PQ}I_{Q,jb},\\
W^{A}_{ij,ab}&=\sum_{P,Q}I_{P,ij}(D_0^{-1})_{PQ}I_{Q,ab},\\
W^{B}_{ib,ja}&=\sum_{P,Q}I_{P,ib}(D_0^{-1})_{PQ}I_{Q,ja},\qquad D_0=J-Q(0) .
\end{aligned}
\end{equation}

\subsection{\raggedright\normalsize Analytic gradient of the AC-$GW$ quasiparticle energy}

\paragraph{The form of the gradient.}
The QP energy is defined implicitly by Eq.~(1), so its total derivative follows from the
implicit-function theorem,
\begin{equation}
\begin{aligned}
\frac{\text{d}\omega_n}{\text{d}\xi}&=Z_n\Big[\frac{\text{d}\epsilon_n}{\text{d}\xi}
+\frac{\text{d}\Sigma^x_{nn}}{\text{d}\xi}
-\frac{\text{d}v^{xc}_n}{\text{d}\xi}
+\Big(\frac{\text{d}\Sigma^{c}_{nn}}{\text{d}\xi}\Big)_{\!\omega}\Big],\\
Z_n&=\Big[1-\frac{\partial\Sigma^{c}_{nn}}{\partial\omega}\Big]^{-1},
\end{aligned}
\end{equation}
where the subscript $\omega$ means that the frequency is held at the root.

The fixed-frequency derivative of the correlation self-energy is the only nontrivial
ingredient: $\Sigma^{c}_{nn}$ depends on the geometry only through $I$, $J$ and
$\{\epsilon_p\}$ (the quadrature nodes and weights are fixed, and $W$, the response vectors $y^k$ and
$y^0$, and the polarization $Q$ are functions of $I$ and $J$, with $Q$ depending on the orbital energies through
$\Delta_{ia}$). Its differential is therefore a linear functional of $\text{d}I$,
$\text{d}J$ and $\text{d}\epsilon$. There exist three correlation covectors
$\bar{I}^{(c)}$, $\bar{J}^{(c)}$ and $\bar{\epsilon}^{(c)}$ such that
\begin{equation}
\Big(\frac{\text{d}\Sigma^{c}_{nn}}{\text{d}\xi}\Big)_{\!\omega}
=\sum_{P,p,q}\bar{I}^{(c)}_{P,pq}\frac{\text{d}I_{P,pq}}{\text{d}\xi}
+\sum_{P,Q}\bar{J}^{(c)}_{PQ}\frac{\text{d}J_{PQ}}{\text{d}\xi}
+\sum_p\bar{\epsilon}^{(c)}_{p}\frac{\text{d}\epsilon_p}{\text{d}\xi}.
\end{equation}
Here and below, a covector means an element of the dual space of the primitive
variations: an object whose indices match those of $\delta I$, $\delta J$,
$\delta\epsilon$ or $\delta v^{xc}$, and which contracts linearly with the corresponding
primitive derivative. For example, $\bar{I}^{(c)}_{P,pq}$ is the covector of
$\Sigma^{c}$ with respect to $I_{P,pq}$.
The exact-exchange and exchange--correlation terms of Eq.~(11) are not part of
$\Sigma^{c}$: the former is a functional of $I$ and $J$ and is carried by the exchange
covectors below, while the latter is the single derivative
$-\text{d}v^{xc}_n/\text{d}\xi$. In particular, no separate covector is needed for
$v^{xc}$.

\paragraph{Exact-exchange covectors.}
Let $y^0_{P,pq}\equiv\sum_Q(J^{-1})_{PQ}I_{Q,pq}$. The exact-exchange self-energy is
\begin{equation}
\Sigma^{x}_{nn}=-\sum_i\sum_{P,Q}I_{P,ni}(J^{-1})_{PQ}I_{Q,ni}.
\end{equation}
The exchange part of the static term is $\Sigma^{x}_{nn}$, with the differential
\begin{equation}
\delta\Sigma^{x}_{nn}
=-2\sum_i\sum_P \delta I_{P,ni}\,y^0_{P,ni}
+\sum_i\sum_{P,Q} y^0_{P,ni}\,\delta J_{PQ}\,y^0_{Q,ni},
\end{equation}
which gives the exchange covectors
\begin{equation}
\bar{I}^{(x)}_{P,ni}=-2y^0_{P,ni},\qquad
\bar{J}^{(x)}_{PQ}=\sum_i y^0_{P,ni} y^0_{Q,ni}.
\end{equation}
These are the complete contribution of the exact-exchange part of the static term to the
$I$ and $J$ covectors. They are not included in Eq.~(12) and are added when the full QP
gradient is assembled below.

\paragraph{Correlation covectors at a fixed quadrature node.}Differentiating Eq.~(7) at
fixed $u_k$ and fixed external frequency yields a linear combination of the derivatives of
the screened interaction and of the orbital energies,
\begin{equation}
\begin{aligned}
\Big(\frac{\text{d}\Sigma^{c}_{nn}}{\text{d}\xi}\Big)_{\!\omega} &=\sum_{k}c_{nk}\,\frac{\text{d}W^c_{nn}(iu_k)}{\text{d}\xi}
+\sum_p \bar{\epsilon}^{(t)}_{p}\,\frac{\text{d}\epsilon_p}{\text{d}\xi},\\
c_{nk}&=w_k\,\frac{t_n}{t_n^{2}+u_k^{2}},
\end{aligned}
\end{equation}
where the second term collects the derivative of the kernel $t_n/(t_n^{2}+u_k^{2})$
with respect to the orbital energy through $t_n$ ($\partial t_n/\partial\epsilon_n=-1$).
Since in Eq.(7) only $t_n$ carries the orbital-energy dependence directly,
\begin{equation}
\bar{\epsilon}^{(t)}_{p}=\delta_{pn}\sum_k w_k\,\frac{t_n^{2}-u_k^{2}}
{\big(t_n^{2}+u_k^{2}\big)^{2}}\,W^c_{nn}(iu_k);
\end{equation}
The differential of a screened matrix element follows from
\begin{equation}
\begin{aligned}
\delta W^c_{nm}(iu_k) &=2\sum_P \delta I_{P,nm}\big(y^{k}_{P,nm}-y^0_{P,nm}\big)\\
&\quad-\sum_{P,Q} y^{k}_{P,nm}\big(\delta J_{PQ}-\delta Q_{PQ}(iu_k)\big)y^{k}_{Q,nm}\\
&\quad+\sum_{P,Q} y^0_{P,nm}\,\delta J_{PQ}\,y^0_{Q,nm},
\end{aligned}
\end{equation}
with
\begin{equation}
\begin{aligned}
y^{k}_{P,nm}&\equiv\sum_Q\big[(J-Q(iu_k))^{-1}\big]_{PQ}I_{Q,nm},\\
y^0_{P,nm}&\equiv\sum_Q(J^{-1})_{PQ}I_{Q,nm},
\end{aligned}
\end{equation}
and the polarization response is
\begin{equation}
\begin{aligned}
\delta Q_{PQ}(iu_k)&=\sum_{ia}\Big[\delta I_{P,ia}I_{Q,ia}+I_{P,ia}\delta I_{Q,ia}\Big]d_{ia}(iu_k)\\
&\quad+I_{P,ia}I_{Q,ia}\,d'_{ia}(iu_k)\,\delta\Delta_{ia},
\end{aligned}
\end{equation}
with $d'_{ia}=\partial d_{ia}/\partial\Delta_{ia}$ and
$\delta\Delta_{ia}=\delta\epsilon_a-\delta\epsilon_i$. Inserting this $\delta Q$ into the
differential of $W$, and the resulting $\delta W$ into Eq.~(16), reading the
coefficients of $\delta I$, $\delta J$ and $\delta\Delta$ gives the correlation covectors
of the AC-$GW$ gradient,
\begin{equation}
\bar{I}^{(W)}_{P,pq}=2\,\delta_{pn}\sum_k c_{nk}\big(y^{k}_{P,nq}-y^0_{P,nq}\big),
\end{equation}
\begin{equation}
\begin{aligned}
\bar{I}^{(Q)}_{P,ia}&=2\sum_k d_{ia}(iu_k)\sum_Q V_{k,PQ}\,I_{Q,ia},\\
V_{k,PQ}&=\sum_m c_{mk}\,y^{k}_{P,nm} y^{k}_{Q,nm},
\end{aligned}
\end{equation}
\begin{equation}
\bar{J}^{(c)}_{PQ}=\sum_{m,k}c_{mk}\Big(y^0_{P,nm}y^0_{Q,nm}
-y^{k}_{P,nm}y^{k}_{Q,nm}\Big),
\end{equation}
where Eq.~(21) comes directly from Eq.~(18), and Eqs.(22)(23) come from inserting Eq.(20) into Eq.(18) and reading the coefficient of $\delta I_{P,ia}$ and $I_{Q,ia}$. The prefactors $c_{nk}$ and $c_{mk}$ come from further inserting Eq.(18) into Eq.(16), and the prefactor $2$ in Eq.~(21) and Eq.~(22) comes from the symmetry of $I$.
The polarization channel additionally produces
the orbital-energy covector
\begin{equation}
\begin{aligned}
\bar{\epsilon}^{(\Delta)}_{p}&=\sum_{ia}\frac{\partial\Delta_{ia}}{\partial\epsilon_p}\,g_{ia},\\
g_{ia}&=\sum_k\sum_{P,Q} I_{P,ia}\,V_{k,PQ}\,I_{Q,ia}\,d'_{ia}(iu_k).
\end{aligned}
\end{equation}
The correlation covectors in Eq.~(12) are therefore
\begin{equation}
\bar{I}^{(c)}_{P,pq}=\bar{I}^{(W)}_{P,pq}+\bar{I}^{(Q)}_{P,pq},\qquad
\bar{\epsilon}^{(c)}_{p}=\bar{\epsilon}^{(t)}_{p}+\bar{\epsilon}^{(\Delta)}_{p},
\end{equation}
with $\bar{J}^{(c)}$ the metric covector of Eq.~(23). Adding the exchange covectors of
Eq.~(15) and the explicit $\epsilon_n$ term gives the full GW covectors
\begin{equation}
\begin{aligned}
\bar{I}^{\rm GW}_{P,pq}&=\bar{I}^{(c)}_{P,pq}+\bar{I}^{(x)}_{P,pq},\\
\bar{J}^{\rm GW}_{PQ}&=\bar{J}^{(c)}_{PQ}+\bar{J}^{(x)}_{PQ},\\
\bar{\epsilon}^{\rm GW}_{p}&=\delta_{pn}+\bar{\epsilon}^{(c)}_{p},
\end{aligned}
\end{equation}
and the full QP gradient
\begin{equation}
\begin{aligned}
\frac{\text{d}\omega_n}{\text{d}\xi} &=Z_n\Big[\sum_{P,p,q}\bar{I}^{\rm GW}_{P,pq}\frac{\text{d}I_{P,pq}}{\text{d}\xi}
+\sum_{P,Q}\bar{J}^{\rm GW}_{PQ}\frac{\text{d}J_{PQ}}{\text{d}\xi}\\
&\quad+\sum_p\bar{\epsilon}^{\rm GW}_{p}\frac{\text{d}\epsilon_p}{\text{d}\xi}
-\frac{\text{d}v^{xc}_n}{\text{d}\xi}\Big].
\end{aligned}
\end{equation}
The last term is the direct derivative of the exchange--correlation potential and carries
no covector. The correlation covectors have exactly the same structure as in the CD
case: the only difference is that in CD it is also required to evaluate covectors for real frequencies and to include a derivative with respect of the real frequency.

\paragraph{Differentiation of the numerical continuation.}
The remaining ingredient is the derivative of the Pad\'e map $\mathcal{P}$ of Eq.~(6).
The approximant is a continued fraction in the evaluation point $z=\omega+i\eta_{\rm ac}$
whose coefficients are fixed by the samples $S_k\equiv\Sigma^{c}_{nn}(\xi_k)$, taken on the
Fermi-shifted imaginary axis,
\begin{equation}
\mathcal{P}(z)=a_0+\frac{z-\xi_0}{a_1+\dfrac{z-\xi_1}{a_2+\cdots}},
\qquad \xi_k=e_F+i\lambda_k,
\end{equation}
and its response to a change of geometry is, by the chain rule,
\begin{equation}
\Big(\frac{\text{d}\Sigma^{c}_{nn}}{\text{d}\xi}\Big)_{\!\omega}
=\sum_k\frac{\partial\mathcal{P}}{\partial S_k}\Big|_{\omega}\,\frac{\text{d}S_k}{\text{d}\xi}
+\frac{\partial\mathcal{P}}{\partial e_F}\Big|_{\omega}\,\frac{\text{d}e_F}{\text{d}\xi},
\end{equation}
where the Jacobian $\partial\mathcal{P}/\partial S_k$ is evaluated once per geometry (it
involves only the continued-fraction coefficients and the root $\omega_n$), the last term
collects the explicit dependence of the approximant on the Fermi shift, which moves every
node $\xi_k$, and the sample derivative
\begin{equation}
\frac{\text{d}S_k}{\text{d}\xi}=\sum_{P,p,q}\bar{I}^{(k)}_{P,pq}\frac{\text{d}I_{P,pq}}{\text{d}\xi}
+\sum_{P,Q}\bar{J}^{(k)}_{PQ}\frac{\text{d}J_{PQ}}{\text{d}\xi}
+\sum_p\bar{\epsilon}^{(k)}_{p}\frac{\text{d}\epsilon_p}{\text{d}\xi}
\end{equation}
is exactly the primitive contraction of the correlation covectors in Eqs.~(16)--(24) with
the node coefficient of Eq.~(16) replaced by the sample kernel at $\xi_k$. Because the
Pad\'e coefficients are determined by all samples simultaneously, the
Jacobian is a
$n_{\rm sample}\times n_{\rm sample}$ object. Equation~(29) shows that for AC-$GW$, the gradient of the correlation self-energy needs to be differentiated only at the $n_{\rm sample}$ imaginary quadrature nodes, where $n_{\rm sample}$ is independent from the size of the system or the orbital of interest, making the AC gradient more affordable compared to the CD gradient for large systems. The full expressions of the Jacobian are collected in the
Supporting Information (Sec.~S1).

\paragraph{Relation to the CD and CD-WAC variants.}
The CD self-energy \cite{schindlmayr1998,golze2018} differs from Eq.~(7) by an
additional sum over the residues corresponding to poles of the Green's function enclosed by the contour
deformation,
\begin{equation}
\begin{aligned}
\Sigma^{c,\rm CD}_{nn}(\omega)&=\Sigma^{c,\rm AC}_{nn}(\omega)
+\sum_m s_m\,W^c{nm}(\zeta_m),\\
\zeta_m&=|\epsilon_m-\omega|,
\end{aligned}
\end{equation}
with the pole weights $s_m=\pm1$ selected relative to the Fermi level. The residue term
is not continued and therefore keeps the correct behaviour of the self-energy near and
below the Fermi level, which is why CD describes deep and core levels better than AC; it also
means that a CD gradient requires $W$ at every residue frequency $\zeta_m$, the number of which equals as the number of orbitals between the orbital of interest and the Fermi level, and often becomes very large when the orbital is very far from the Fermi level or when the system is very large. The CD-WAC variant \cite{panades2023} avoids evaluating the exact CD real-axis self-energy with a numerical fit---the fit has sampling frequencies both on the imaginary-axis and on the real-axis---so that the residue sum is replaced by a fit whose cost is, like in AC, independent of
the number of residues, while the accuracy of CD for core levels is retained. Both the gradients of the
CD the CD-WAC variants are derived in the Supporting
Information (Secs.~S2-6).

\subsection{\raggedright\normalsize Analytic gradient of the BSE excitation energy}

The BSE excitation energy depends on the geometry through the QP energies that dress the
diagonal of $A$ and through the primitive objects that enter the kernel. The
amplitudes contribute no derivative of their own, because they are the stationary
eigenvectors of Eq.~(8); hence
\begin{equation}
\begin{aligned}
\frac{\text{d}\Omega_S}{\text{d}\xi}&=\sum_p c_p^{(S)}\frac{\text{d}E^{GW}_p}{\text{d}\xi}
+\Big(\frac{\text{d}\Omega_S^{\rm K}}{\text{d}\xi}\Big)_{\!X,Y},\\
c_i^{(S)}&=-\sum_a\big(X_{ia}^{2}+Y_{ia}^{2}\big),\quad
c_a^{(S)}=+\sum_i\big(X_{ia}^{2}+Y_{ia}^{2}\big).
\end{aligned}
\end{equation}
The first term is the amplitude-weighted sum of the full QP gradients constructed above, so the
whole machinery of the preceding subsection enters through back-propagation with the
amplitude weight row $c^{(S)}$; the renormalization factors $Z_p$ are already contained in
that back-propagation. The
second term is a Hellmann--Feynman expectation value at fixed amplitudes,
\begin{equation}
\begin{aligned}
\Big(\frac{\text{d}\Omega_S^{\rm K}}{\text{d}\xi}\Big)_{\!X,Y} &=X^{T}\text{d}A_K\,X+Y^{T}\text{d}A_K\,Y\\
&\quad+X^{T}\text{d}B_K\,Y+Y^{T}\text{d}B_K\,X,\\
A_K&=\kappa v-W^{A},\quad B_K=\kappa v-W^{B},
\end{aligned}
\end{equation}
a linear functional of the kernel differentials. The kernel depends on $I$ through three
channels: the bare Coulomb term $v=I^{T}J^{-1}I$, the two $I$ factors in the screened
exchange $W=I^{T}D_0^{-1}I$, and the $D_0$ dependence of $W$. We now give the
corresponding covectors.

For the bare Coulomb term, define
\begin{equation}
\begin{aligned}
v_{P,ia}&\equiv\sum_Q(J^{-1})_{PQ}I_{Q,ia},\\
vz_P&\equiv\sum_{ia} v_{P,ia} z_{ia},\\
z_{ia}&=X_{ia}+Y_{ia},
\end{aligned}
\end{equation}
so that
\begin{equation}
\bar{I}^{\rm v}_{P,ia}=2\kappa\,z_{ia}\,vz_P,\qquad
\bar{J}^{\rm v}_{PQ}=-\kappa\,vz_P vz_Q.
\end{equation}
For the two $I$ factors of the screened exchange term, with
\begin{equation}
u_{P,ia}\equiv\sum_Q(D_0^{-1})_{PQ}I_{Q,ia},
\end{equation}
the amplitude contractions give
\begin{equation}
\begin{aligned}
\bar{I}^{\rm W,A}_{P,ij} &=-\sum_{ab}\big(X_{ia}u_{P,ab}X_{jb}+Y_{ia}u_{P,ab}Y_{jb}\big),\\
\bar{I}^{\rm W,A}_{P,ab} &=-\sum_{ij}\big(X_{ia}u_{P,ij}X_{jb}+Y_{ia}u_{P,ij}Y_{jb}\big),\\
\bar{I}^{\rm W,B}_{P,ia} &=-2\sum_{jk}\big(X_{ik}u_{P,jk}Y_{ja}+Y_{ik}u_{P,jk}X_{ja}\big).
\end{aligned}
\end{equation}
The $D_0$ dependence is obtained from the matrix identity
\begin{equation}
I^{T}\delta(D_0^{-1})I
=-I^{T}D_0^{-1}\delta D_0\,D_0^{-1}I,
\end{equation}
whose left- and right-hand sides are products of two $u=D_0^{-1}I$ tensors.
The overall minus sign from $-W^{A}$ and $-W^{B}$ in Eq.~(33) cancels the minus of the
identity. Contracting these products with the amplitudes $X$ and $Y$ as in Eq.~(33) gives the
$D_0$-covectors
\begin{equation}
\begin{aligned}
\bar{D}^{A}_{PQ} &=\sum_{ijab}\big(X_{ia}X_{jb}+Y_{ia}Y_{jb}\big)u_{P,ij}u_{Q,ab},\\
\bar{D}^{B}_{PQ} &=\sum_{iajk}\big(X_{ik}Y_{ja}+Y_{ik}X_{ja}\big)u_{P,ia}u_{Q,jk}.
\end{aligned}
\end{equation}
These are covectors with respect to $D_0$; they are not
needed explicitly in the final gradient, but they are the intermediate covectors that
carry the $D_0$ dependence through $Q_0$ below.
Chaining the differential of the static response
\begin{equation}
\begin{aligned}
\delta Q_{0,PQ}&=\sum_{ia}\big[(\delta I_{P,ia})I_{Q,ia}+I_{P,ia}(\delta I_{Q,ia})\big]d_{ia}(0)\\
&\quad+I_{P,ia}I_{Q,ia}d'_{ia}(0)\,\delta\Delta_{ia}
\end{aligned}
\end{equation}
through $\bar{D}_{PQ}=\bar{D}^{A}_{PQ}+\bar{D}^{B}_{PQ}$ completes the covectors,
\begin{equation}
\begin{aligned}
\bar{I}^{\rm D0}_{P,ia} &=-2\,d_{ia}(0)\sum_Q\bar{D}_{PQ}I_{Q,ia}
=\frac{8}{\Delta_{ia}}\sum_Q\bar{D}_{PQ}I_{Q,ia},\\
\bar{\epsilon}^{\rm D0}_{p}&=
\begin{cases}
-\sum_a\dfrac{4}{\Delta_{pa}^{2}}\sum_{P,Q}I_{P,pa}\bar{D}_{PQ}I_{Q,pa}, & p\ \text{occ.},\\[2pt]
+\sum_i\dfrac{4}{\Delta_{ip}^{2}}\sum_{P,Q}I_{P,ip}\bar{D}_{PQ}I_{Q,ip}, & p\ \text{vir.},
\end{cases}
\end{aligned}
\end{equation}
\begin{equation}
\bar{J}^{\rm K}_{PQ}=\bar{J}^{\rm v}_{PQ}+\mathrm{sym}\big(\bar{D}^{A}+\bar{D}^{B}\big)_{PQ},
\end{equation}
so that the BSE gradient has the same contraction structure as the full QP gradient above,
\begin{equation}
\begin{aligned}
\frac{\text{d}\Omega_S}{\text{d}\xi} &=\sum_{P,p,q}\bar{I}^{\rm BSE}_{P,pq}\frac{\text{d}I_{P,pq}}{\text{d}\xi}
+\sum_{P,Q}\bar{J}^{\rm BSE}_{PQ}\frac{\text{d}J_{PQ}}{\text{d}\xi}\\
&\quad+\sum_p\bar{\epsilon}^{\rm BSE}_{p}\frac{\text{d}\epsilon_p}{\text{d}\xi}
-\sum_p c_p^{(S)}\frac{\text{d}v^{xc}_p}{\text{d}\xi},
\end{aligned}
\end{equation}
with $\bar{I}^{\rm BSE}_{P,pq}=\bar{I}^{\rm GW}_{P,pq}[c^{(S)}]+\bar{I}^{\rm K}_{P,pq}$,
$\bar{J}^{\rm BSE}_{PQ}=\bar{J}^{\rm GW}_{PQ}[c^{(S)}]+\bar{J}^{\rm K}_{PQ}$ and
$\bar{\epsilon}^{\rm BSE}_{p}=\bar{\epsilon}^{\rm GW}_{p}[c^{(S)}]+\bar{\epsilon}^{\rm D0}_{p}$;
the last term is the back-propagation of the QP exchange--correlation derivative with the
amplitude weight row $c^{(S)}$. The exact-exchange derivative is already contained in these total
covectors through $\bar{I}^{\rm GW}=\bar{I}^{(c)}+\bar{I}^{(x)}$ and
$\bar{J}^{\rm GW}=\bar{J}^{(c)}+\bar{J}^{(x)}$. Here $\bar{I}^{\rm K}_{P,pq}$ collects the bare-Coulomb, screened-exchange and
$D_0$-induced kernel covectors with auxiliary index $P$ and pair $pq$. The kernel changes only the
covectors, never the way in which they are contracted with the primitive gradients.
This is why the BSE gradient can be built on the AC, CD or CD-WAC variant of Sec.~2.2
without any additional derivation: one only needs the QP covectors of that variant,
evaluated with the amplitude weight row $c^{(S)}$.

\subsection{\raggedright\normalsize Gradients of the primitives}

It remains to evaluate the primitive gradients and the $v^{xc}$ derivative, which are the
same for all $GW$ variants and for the BSE. The exact-exchange covectors of Eq.~(15) and
the $v^{xc}$ derivative below are the two static contributions to the QP gradient. Writing the first-order MO
coefficients as $C^{\xi}=CU^{\xi}$,
orthonormality fixes the symmetric part of the rotation and the differentiated Roothaan
condition the antisymmetric part,
\begin{equation}
\begin{aligned}
U^{\xi}_{pp}&=-\tfrac12\bar S^{\xi}_{pp},\\
U^{\xi}_{pq}&=\frac{\bar F^{\xi}_{pq}-\epsilon_q\bar S^{\xi}_{pq}}{\epsilon_q-\epsilon_p}
\quad(p\neq q),\\
\epsilon_p^{\xi}&=\bar F^{\xi}_{pp}-\epsilon_p\bar S^{\xi}_{pp},
\end{aligned}
\end{equation}
\begin{equation}
\begin{aligned}
\bar F^{\xi}&=C^{T}F^{\xi,\rm ao}C,\\
F^{\xi,\rm ao}&=h^{\xi}+J^{\xi}[D]-\tfrac12 c_xK^{\xi}[D]\\
&\quad+v_{xc}^{\xi}[\rho]+f_{xc}[\rho^{\xi}],\\
D^{\xi}&=\sum_i\big(C_i^{\xi}C_i^{T}+C_iC_i^{\xi,T}\big),
\end{aligned}
\end{equation}
where the first three terms are the skeleton Fock derivative, the fourth the explicit
exchange--correlation derivative at frozen density and orbitals, and the last the
relaxation through the first-order density, whose occupied--virtual block is determined
by the coupled-perturbed SCF equation. Because every gradient contribution is linear in
$U^{\xi}$, a single adjoint (Z-vector) equation replaces one response per Cartesian
component,
\begin{equation}
\begin{aligned}
(A+B)^{T}Z&=L,\\
L_{pi}&=4\big(C^{T}V_{\rm til}C_{\rm occ}\big)_{pi},\\
V_{\rm til}&=J[\Lambda]-\tfrac12c_xK[\Lambda]+f_{xc}[\Lambda],
\end{aligned}
\end{equation}
with $\Lambda$ assembled from the covectors of the preceding subsections. The
three-centre tensor differentiates as
\begin{equation}
\frac{\text{d}I_{P,pq}}{\text{d}\xi}=C^{T}I_P^{\xi}C+(U^{\xi})^{T}I_P+I_PU^{\xi},
\end{equation}
where $I_P^{\xi}$ is the explicit nuclear derivative of the AO three-centre integral
$I_{\mu\nu P}$, while $I_P$ denotes the MO-transformed tensor of Eq.~(3). Contracting
with a covector $\mathsf{Q}$ gives, after folding the orbital-independent
factors once per geometry,
\begin{equation}
\begin{aligned}
\sum_{P,p,q}\mathsf{Q}_{P,pq}\frac{\text{d}I_{P,pq}}{\text{d}\xi} &=\sum_P\big\langle C\,\mathsf{Q}_P\,C^{T},I_P^{\xi}\big\rangle
+2\,\mathrm{Tr}\big[(U^{\xi})^{T}W\big],\\
W&=\sum_P I_P\mathsf{Q}_P,
\end{aligned}
\end{equation}
which replaces two MO transformations of the AO derivative $I^{\xi}$ per perturbation
by one AO-basis contraction. The metric derivative follows from the two-centre integral derivative,
\begin{equation}
\frac{\text{d}J_{PQ}}{\text{d}\xi}=\sum_A\frac{\partial J_{PQ}}{\partial R_A}\cdot\frac{\text{d}R_A}{\text{d}\xi},
\end{equation}
and the diagonal exchange--correlation potential differentiates as
\begin{equation}
\frac{\text{d}v^{xc}_p}{\text{d}\xi}=\Big[\big\langle p\big|\dot v_{xc}\big|p\big\rangle
+2\sum_qU_{qp}v_{xc,qp}+\big\langle p\big|f_{xc}[\rho^{\xi}]\big|p\big\rangle\Big].
\end{equation}
The only place where basis functions are differentiated is thus the AO integral
derivatives $I_P^{\xi}$ and $J^{\xi}$;
because the energy is expressed in the raw (non-orthogonalized) RI representation, no
derivative of $J^{-1/2}$ appears anywhere. Collecting Eqs.~(46)--(48), the AC-$GW$ gradient above and the BSE gradient of
Eq.~(43) are evaluated by the identical machinery: the $GW$ variant and the BSE kernel
change only the total covectors $\bar{I}^{\rm BSE}$, $\bar{J}^{\rm BSE}$ and
$\bar{\epsilon}^{\rm BSE}$, never the way in which they are contracted. The algorithmic consequence is that the adjoint
equation (46), the folding of Eq.~(48) and the explicit integral derivatives are
independent of the target state and are reused for every QP energy and every BSE root.

\section{\raggedright\large Methods}

\paragraph{Functionals and basis sets.}
All calculations use the REST program \cite{rest2025,restgwbse2026}. The source code of REST is available at \url{https://gitee.com/restgroup/rest}. The SCF is
performed with the RI-V approximation \cite{vahtras1993,eichkorn1995} and a spherical
Gaussian basis \cite{weigend2005} with its Coulomb-fitting auxiliary set \cite{weigend2006}; the $GW$ and BSE
steps use def2-SVP for the finite-difference tests and for the DMABN relaxation
and dynamics, def2-TZVPP for the CO minima and def2-TZVP for 1-phenyl-3-(dicyanovinyl)piperidine(abbreviated as DA1 in the rest of this paper), with the corresponding
universal or dedicated RI auxiliary sets. Three
reference functionals are used for the $GW$ gradient validations (all of which are known
starting points for $G_0W_0$ \cite{bruneval2013,vansetten2015,bois2017}): Hartree-Fock,
PBE and PBE0. The geometry optimizations use both PBE0 and PBEh (hybrid functional of PBE with an exact-exchange fraction of
$\alpha=0.45$) for CO and DMABN, and PBEh only for DA1.

\paragraph{Finite-difference tests.}
Gradients are validated by central differences of the respective energy (QP energy for
the $GW$ validations, BSE excitation energy for the BSE validations) with a displacement of the $z$
component of the first atom;
$h_{\rm FD}=10^{-4}$~Bohr is used throughout.

\paragraph{Geometry optimizations.}
Excited-state structures are optimized with the AC-$GW$-BSE analytic gradient and the geomeTRIC optimizer\cite{geometric2016} which is intregrated in REST as a module. The active transition space inludes orbitals within a $\pm$2.0~Ha window around the gap for CO and $\pm$0.4~Ha
for DMABN and DA1. The target state is identified by
the overlap of its excitation-density fingerprint with the previous step throughout the optimization procedure, and the
convergence is
$|g|_{\rm rms}<10^{-3}$~Ha/Bohr, $|g|_{\rm max}<3\times10^{-3}$~Ha/Bohr and
$|\Delta E|<10^{-6}$~Ha.

\paragraph{Trajectory surface hopping simulations.}
Trajectory surface-hopping (TSH) simulations are performed with a modified version of
the SHARC 3.0 program \cite{richter2011,sharc2018,sharc2026} that accepts REST outputs through an
interface layer\cite{mausenberger2025}. The source code of this version of SHARC is available at
\url{https://gitee.com/qirui-gao-2025/sharc}. The electronic-structure step is the
AC-$GW$-BSE@PBE0(TDA) gradient of the present work (def2-SVP, AC-$G_0W_0$ with 16
imaginary-axis samples); the analytic gradient works as the excited-state force, and the
inter-state overlaps returned by REST drive the local-diabatization propagation of the
electronic amplitudes. An ensemble of
\num{30} trajectories is started from a Wigner distribution at \SI{0}{K} built from the
analytic ground-state Hessian, each one in S$_2$ at the vertical excitation energy of
\SI{4.85}{eV}. The nuclear time step is \SI{0.5}{fs} with 25 electronic substeps for the integration of the electronic equation-of-motion,
decoherence is included through the energy-based correction with $\tau=\SI{0.1}{fs}$, and
spin-orbit coupling is neglected. The simulations run for a total of \SI{300}{fs},
the populations of the electronic states are monitored
throughout, and the ring-NMe$_2$ twist angles are measured at t=\SI{300}{fs}.

\section{\raggedright\large Results}

\subsection{\raggedright\normalsize Correctness of the $GW$ and BSE gradients: finite-difference validation}

Table~\ref{tab:gwfd} reports the analytic $GW$ quasiparticle gradient of the HOMO (or the
highest occupied orbital inside the window) for the four test molecules and three
reference functionals, compared with central finite differences, for the three $GW$
variants. All 30 combinations agree, with deviations between
\num{2e-7} and \num{6e-5}~Ha/Bohr, confirming that the analytic nuclear gradient is correctly implemented.

\begin{table*}[h]
\centering
\tabfont
\caption{Analytic minus finite-difference $GW$ quasiparticle gradient
($h_{\rm FD}=10^{-4}$~Bohr, $z$ displacement of the first atom, def2-SVP). Units
Ha/Bohr. The three columns are the CD, AC and CD-WAC variants of the self-energy.}
\label{tab:gwfd}
\begin{tabular}{llccc}
\toprule
System & Reference & CD & AC & CD-WAC \\
\midrule
H$_2$O     & HF   & \num{3.03e-6} & \num{2.70e-6} & \num{3.02e-6} \\
H$_2$O     & PBE  & \num{1.48e-6} & \num{5.97e-5} & \num{4.03e-6} \\
H$_2$O     & PBE0 & \num{2.93e-6} & \num{8.80e-7} & \num{2.99e-6} \\
NH$_3$     & HF   & \num{1.11e-6} & \num{1.11e-6} & \num{1.10e-6} \\
NH$_3$     & PBE  & \num{1.67e-6} & \num{1.12e-6} & \num{1.60e-5} \\
NH$_3$     & PBE0 & \num{1.03e-6} & \num{3.50e-7} & \num{1.31e-6} \\
H$_2$CO    & HF   & \num{1.76e-6} & \num{1.85e-6} & \num{1.75e-6} \\
H$_2$CO    & PBE0 & \num{1.02e-6} & \num{1.86e-6} & \num{2.05e-7} \\
C$_2$H$_4$ & HF   & \num{4.05e-6} & \num{4.31e-6} & \num{4.11e-6} \\
C$_2$H$_4$ & PBE  & \num{7.40e-7} & \num{1.50e-5} & \num{8.40e-7} \\
\bottomrule
\end{tabular}
\end{table*}

Table~\ref{tab:cross} shows that the three variants not only reproduce their own finite
differences but also agree with each other to a good extent. For the Hartree--Fock reference the
quasiparticle energies coincide to better than \num{2e-4}~eV and the gradients to
\num{7e-8}--\num{8e-7}~Ha/Bohr.
For the PBE and PBE0/PBEh references the AC and CD-WAC QP energies differ from CD by
at most \num{0.6}~meV, while the gradients differ by
\num{5e-5}--\num{5e-4}~Ha/Bohr for AC and by
\num{4e-8}--\num{4.7e-4}~Ha/Bohr for CD-WAC; the CD-WAC variant is thus one to three
orders of magnitude closer to the CD result than the plain AC variant, which suggests that CD-WAC successfully reproduces the behaviour of CD self-energy through its numerical fitting procedure.

\begin{table*}[h]
\centering
\tabfont
\caption{Cross-comparison of the three $GW$ variants at a common geometry
(def2-SVP). $\Delta E$ is the quasiparticle-energy difference with respect to full CD
(meV); the last two columns are the largest componentwise difference of the analytic
gradients (Ha/Bohr).}
\label{tab:cross}
\begin{tabular}{llcccc}
\toprule
System & Reference & $\Delta E$(AC$-$CD) & $\Delta E$(WAC$-$CD)
& $\max|g_{\rm CD}-g_{\rm AC}|$ & $\max|g_{\rm CD}-g_{\rm WAC}|$ \\
\midrule
H$_2$O     & HF   & $-0.00$ & $-0.00$ & \num{4.7e-7} & \num{7.0e-8} \\
H$_2$O     & PBE  & $-0.32$ & $+0.00$ & \num{4.8e-4} & \num{1.6e-6} \\
H$_2$O     & PBE0 & $-0.02$ & $+0.00$ & \num{5.5e-5} & \num{4.0e-8} \\
NH$_3$     & HF   & $-0.00$ & $-0.00$ & \num{5.3e-7} & \num{3.0e-7} \\
NH$_3$     & PBE  & $-0.58$ & $+0.00$ & \num{4.8e-4} & \num{5.5e-7} \\
NH$_3$     & PBE0 & $-0.09$ & $+0.00$ & \num{4.9e-4} & \num{3.2e-7} \\
H$_2$CO    & HF   & $-0.00$ & $-0.00$ & \num{3.8e-6} & \num{1.1e-7} \\
H$_2$CO    & PBE0 & $-0.00$ & $+0.00$ & \num{9.9e-6} & \num{1.2e-6} \\
C$_2$H$_4$ & HF   & $+0.00$ & $+0.08$ & \num{3.3e-6} & \num{2.3e-4} \\
C$_2$H$_4$ & PBE  & $+0.12$ & $+0.02$ & \num{2.5e-3} & \num{4.7e-4} \\
\bottomrule
\end{tabular}
\end{table*}

Table~\ref{tab:bsefd} validates the BSE gradient built on the AC-GW self-energy. The
lowest singlet excitation energy and its analytic gradient are compared with central
differences of $\Omega$ for the same four molecules with the PBE0 reference; the
agreement, \num{6e-8}--\num{2e-6}~Ha/Bohr, is as good as that of the underlying $GW$
gradient. Since the $GW$ analytic gradients of all variants have been shown to be equally consistent with their numerical gradients, it is assumed that CD and CD-WAC gradients work equally well for BSE.

\begin{table*}[h]
\centering
\tabfont
\caption{Analytic versus finite-difference gradient of the lowest singlet BSE excitation
energy from AC-GW (PBE0/def2-SVP, $h_{\rm FD}=10^{-4}$~Bohr). Units Ha/Bohr;
$\Omega$ in eV.}
\label{tab:bsefd}
\begin{tabular}{lcccc}
\toprule
System & $\Omega$ (eV) & $|g|$ & $\max|g_{\rm ana}-g_{\rm FD}|$ & relative \\
\midrule
H$_2$O     & 7.6361 & \num{1.430e-1} & \num{9.05e-7} & \num{6.3e-6} \\
NH$_3$     & 7.1580 & \num{7.490e-2} & \num{1.75e-6} & \num{2.3e-5} \\
H$_2$CO    & 3.7104 & \num{1.653e-1} & \num{1.55e-7} & \num{9.4e-7} \\
C$_2$H$_4$ & 8.7192 & \num{8.128e-2} & \num{6.00e-8} & \num{7.4e-7} \\
\bottomrule
\end{tabular}
\end{table*}

\subsection{\raggedright\normalsize Excited-state geometry optimization}

\paragraph{CO.}
Table~\ref{tab:exc} collects the AC-$GW$-BSE minima of the $A^1\Pi$ and $a^3\Pi$ states of
CO, obtained with the PBE0 and PBEh starting points, the def2-TZVPP basis set and
the Tamm--Dancoff approximation, and compares them with experimental ground state and excited state geometries\cite{nist_co_diatomic}. The $A^1\Pi$ bond length is \num{0.115} and \num{0.101}{\AA} longer
than in the ground state for the two references, and reproduces the experimental value
\cite{nist_co_diatomic} to \num{0.008} (PBE0) and \num{0.006}{\AA} (PBEh). For $a^3\Pi$ the PBE0 bond length is \num{0.011}{\AA} shorter than the
experimental one, while the PBEh reference gives \num{0.023}{\AA} in both
comparisons. This shows that optimizing excited-state geometries with the AC-$GW$-BSE analytic gradient gives quantitatively close results compared to experimental values for such simple model systems.

\begin{table*}[h]
\centering
\tabfont
\caption{Excited-state minima of CO from the AC-$GW$-BSE(TDA) analytic gradient
 with the PBE0 and
PBEh (both under def2-TZVPP) as starting points, compared with the ground-state and experimental
excited-state bond lengths of Ref.~\cite{nist_co_diatomic}. Bond lengths in \AA,
excitation energies in eV.}
\label{tab:exc}
\begin{tabular}{llcccccc}
\toprule
System & State & Quantity & S$_0$ & PBE0 & PBEh & Expt. \\
\midrule
CO & $A^1\Pi$ & $R$(C-O)& 1.128 & 1.243 & 1.229 & 1.235 \\
CO & $a^3\Pi$ & $R$(C-O)& 1.128 & 1.195 & 1.183 & 1.206 \\
\bottomrule
\end{tabular}
\\[3pt]
{\scriptsize The S$_0$ and Expt.\ bond lengths are from Ref.~\cite{nist_co_diatomic};
the S$_0$ column gives the ground-state reference geometry from which the optimization
starts. The present structures are converged to $|g|\le2\times10^{-3}$~Ha/Bohr.}
\end{table*}

\paragraph{DA1.}

\begin{figure}[t]
\centering
\includegraphics[width=0.675\linewidth]{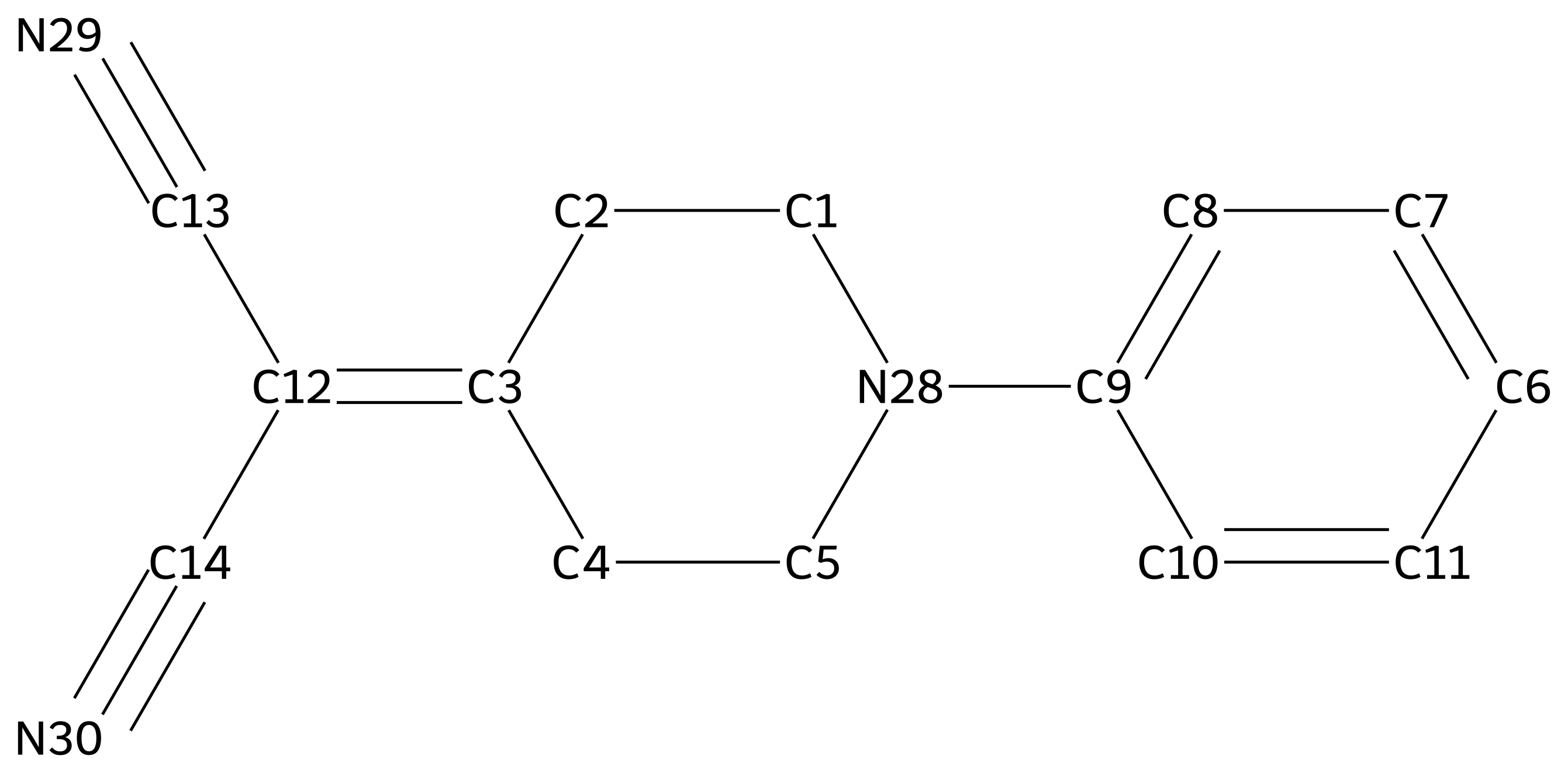}
\caption{Structure of DA1 (1-phenyl-3-(dicyanovinyl)piperidine) with all
non-hydrogen atoms numbered according to the convention of Ref.~\cite{tapavicza2009}.}
\label{fig:da1_structure}
\end{figure}

A second application is the charge-transfer state of
1-phenyl-3-(dicyanovinyl)piperidine (DA1), the prototype $\sigma$-bridged
donor--acceptor molecule whose excited-state properties and dynamics were studied by
Tapavicza \emph{et al.} \cite{tapavicza2009} and for which CC2 and TD-PBE0 reference
values are available. The structure is shown in Fig.~\ref{fig:da1_structure}. We
relaxed the charge-transfer minimum with the $GW$-BSE@PBEh(TDA)
gradient (under def2-TZVP). The optimized geometry is reported in the supporting materials.

The signed pyramidalization angles are defined as
\begin{equation}
\label{eq:pyr}
\begin{aligned}
\theta_{\rm pyr}({\rm N28}) &= \frac{1}{3}\sum_{i=1}^{3}
  \operatorname{wrap}_{180}\!\left[180^\circ-\phi_i^{N28}\right],\\
\theta_{\rm pyr}({\rm C3}) &= \frac{1}{3}\sum_{i=1}^{3}
  \operatorname{wrap}_{180}\!\left[\phi_i^{C3}-180^\circ\right],
\end{aligned}
\end{equation}
where
$\phi_1^{N28}=D({\rm C1},{\rm C5},{\rm C9},{\rm N28})$,
$\phi_2^{N28}=D({\rm C5},{\rm C9},{\rm C1},{\rm N28})$,
$\phi_3^{N28}=D({\rm C9},{\rm C1},{\rm C5},{\rm N28})$,
and
$\phi_1^{C3}=D({\rm C2},{\rm C12},{\rm C4},{\rm C3})$,
$\phi_2^{C3}=D({\rm C12},{\rm C4},{\rm C2},{\rm C3})$,
$\phi_3^{C3}=D({\rm C4},{\rm C2},{\rm C12},{\rm C3})$,
with $D(a,b,c,d)$ the signed dihedral in $(-180^\circ,180^\circ]$ and
$\operatorname{wrap}_{180}(x)=((x+180^\circ)\bmod 360^\circ)-180^\circ$.
The N28 and C3 definitions carry opposite signs because the atom ordering in
the two reference dihedral sets is reversed.

\begin{figure}[t]
\centering
\includegraphics[width=0.882\linewidth]{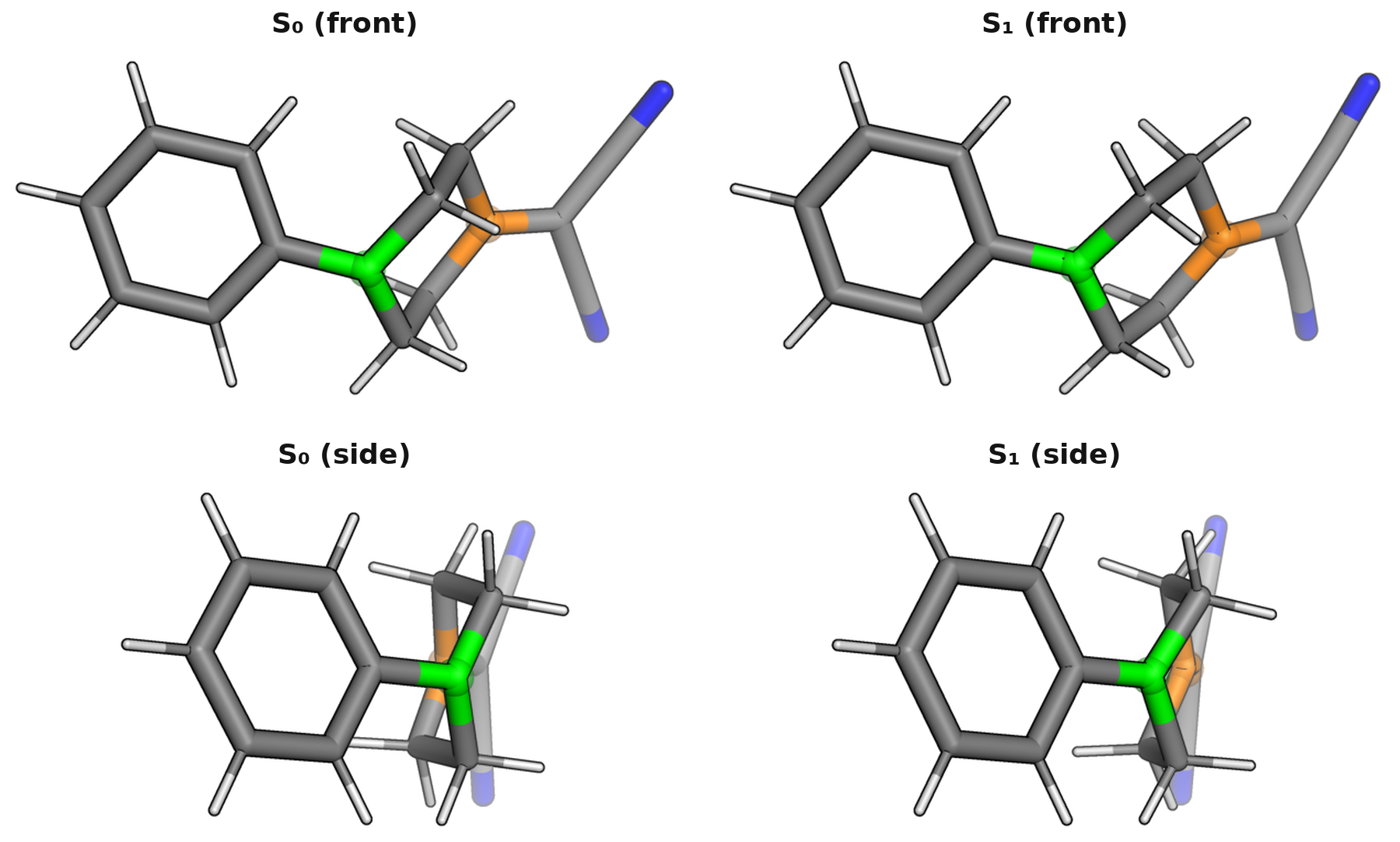}
\caption{Ground- and excited-state geometries of DA1. Top: front views of the
$GW$-BSE@PBEh(TDA) ground-state minimum (left) and the $GW$-BSE@PBEh(TDA) S$_1$ minimum (right). Bottom: the corresponding side views, obtained by
rotating the viewing direction towards the ring plane. The orange sphere marks the
acceptor carbon C3 and the green sphere the amine nitrogen N28.}
\label{fig:da1_3d}
\end{figure}

The converged structure is compared with the literature values in
Table~\ref{tab:da1}. The minimum remains twisted ($\phi=22.4^\circ$), lengthens the
N-C$_{\rm ipso}$ bond by \SI{0.099}{\AA} with respect to the ground state
($1.245$~\AA), and places the excitation at \SI{2.858}{eV}. The amine nitrogen
pyramidalizes by $+7.1^\circ$, i.e. in the same direction as CC2 ($+10.2^\circ$),
whereas TD-PBE0 gives $-4.8^\circ$, the opposite direction and fails to characterize this behaviour. Meanwhile, the $C_3$ carbon
pyramidalizes by $-16.0^\circ$, i.e. in the same out-of-plane direction as CC2
($-22.7^\circ$) and TD-PBE0 ($-21.3^\circ$) and with a comparable magnitude, thus preserving key behaviour observed in CC2 and TD-PBE0. The
present PBEh/TDA structure therefore reproduces both the sizable $C_3$
pyramidalization and the sign of the N28 pyramidalization of CC2, removing the
opposite N28 angle obtained with TD-PBE0. The structural changes are illustrated
in Fig.~\ref{fig:da1_3d}, where the top and bottom rows show front and side views,
respectively.

\begin{table*}[h]
\centering
\tabfont
\caption{Excited-state minimum of DA1 from the AC-GW-BSE analytic gradient
(BSE@AC-$G_0W_0$, def2-TZVP) at the PBEh/TDA level, compared with literature
values. Bond lengths in \AA, angles in degrees, energies in eV.}
\label{tab:da1}
\begin{tabular}{lccc}
\toprule
Quantity & PBEh/TDA & CC2$^{a}$ & TD-PBE0$^{a}$ \\
\midrule
$R$(N28--C9) / \AA      & 1.344 & 1.379 & 1.337 \\
pyr(N28) / $^\circ$     & $+7.1$ & $+10.2$ & $-4.8$ \\
pyr(C3) / $^\circ$   & $-16.0$ & $-22.70$ & $-21.34$ \\
\bottomrule
\end{tabular}
\\[3pt]
{\scriptsize $^{a}$Previously reported CC2 and TD-PBE0 values for the same dye
\cite{tapavicza2009}. The present PBEh/TDA
value has the same sign as CC2 and the opposite sign to TD-PBE0. The $C_3$ carbon
pyramidalizes by $-16.0^\circ$ in PBEh/TDA, compared with $-22.70^\circ$ in CC2
and $-21.34^\circ$ in TD-PBE0. The pyramidalization angles are defined by Eq.~\eqref{eq:pyr}.}
\end{table*}

\paragraph{DMABN.}

\begin{figure}[t]
\centering
\includegraphics[width=0.522\linewidth]{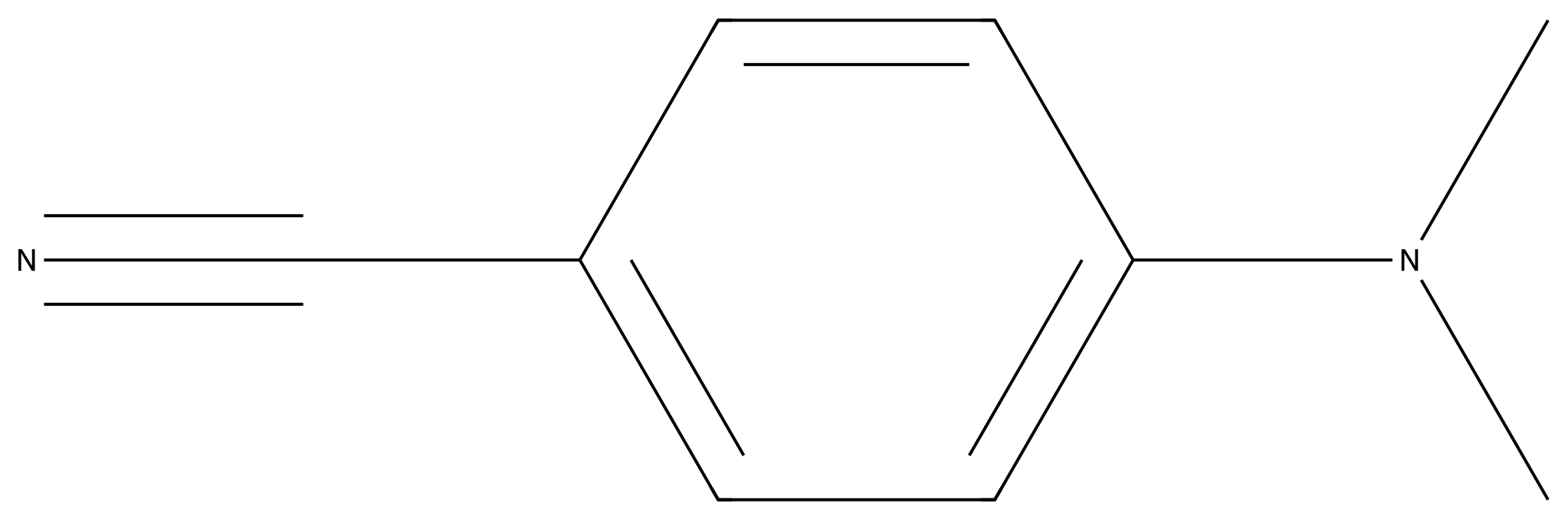}
\caption{Structure of DMABN (4-(dimethylamino)benzonitrile).}
\label{fig:dmabn_structure}
\end{figure}

4-(dimethylamino)benzonitrile (DMABN) is the textbook example of a twisted
intramolecular charge-transfer (TICT) chromophore (Fig.~\ref{fig:dmabn_structure}):
its dual fluorescence has been
interpreted, since the work of Grabowski and co-workers \cite{grabowski2003}, as
emission from a charge-transfer state in which the dimethylamino group is twisted out of
the ring plane. The twist coordinate is therefore a stringent test of an excited-state
gradient, and it has recently been used as a benchmark for machine-learned
nonadiabatic dynamics driven by TD-DFT \cite{mausenberger2025}.

We relaxed the charge-transfer state of DMABN with the $GW$-BSE@PBEh(TDA)
and $GW$-BSE@PBE0(TDA) gradients. The optimized geometry is reported in the supporting materials. Both optimizations converge to a twisted
charge-transfer minimum whose geometry and converging behaviour are impacted by the amount of exact exchange:with PBEh the minimum is reached in eleven steps, the dimethylamino
group is perpendicular to the ring (\SI{90.0}{\degree});with PBE0 the minimum is found after 54 steps at a twist of
\SI{84.4}{\degree}.

In both cases the relaxation lengthens the C$_{\rm ipso}$--N bond
($1.387$ and $1.422$~\AA\ against $1.369$~\AA\ in the ground state) and increases the
ring bond alternation from \num{0.032} to \num{0.072} and \num{0.084}~\AA, i.e.\ the
quinoidal distortion expected for a charge-transfer state.

\begin{figure}[t]
\centering
\includegraphics[width=0.882\linewidth]{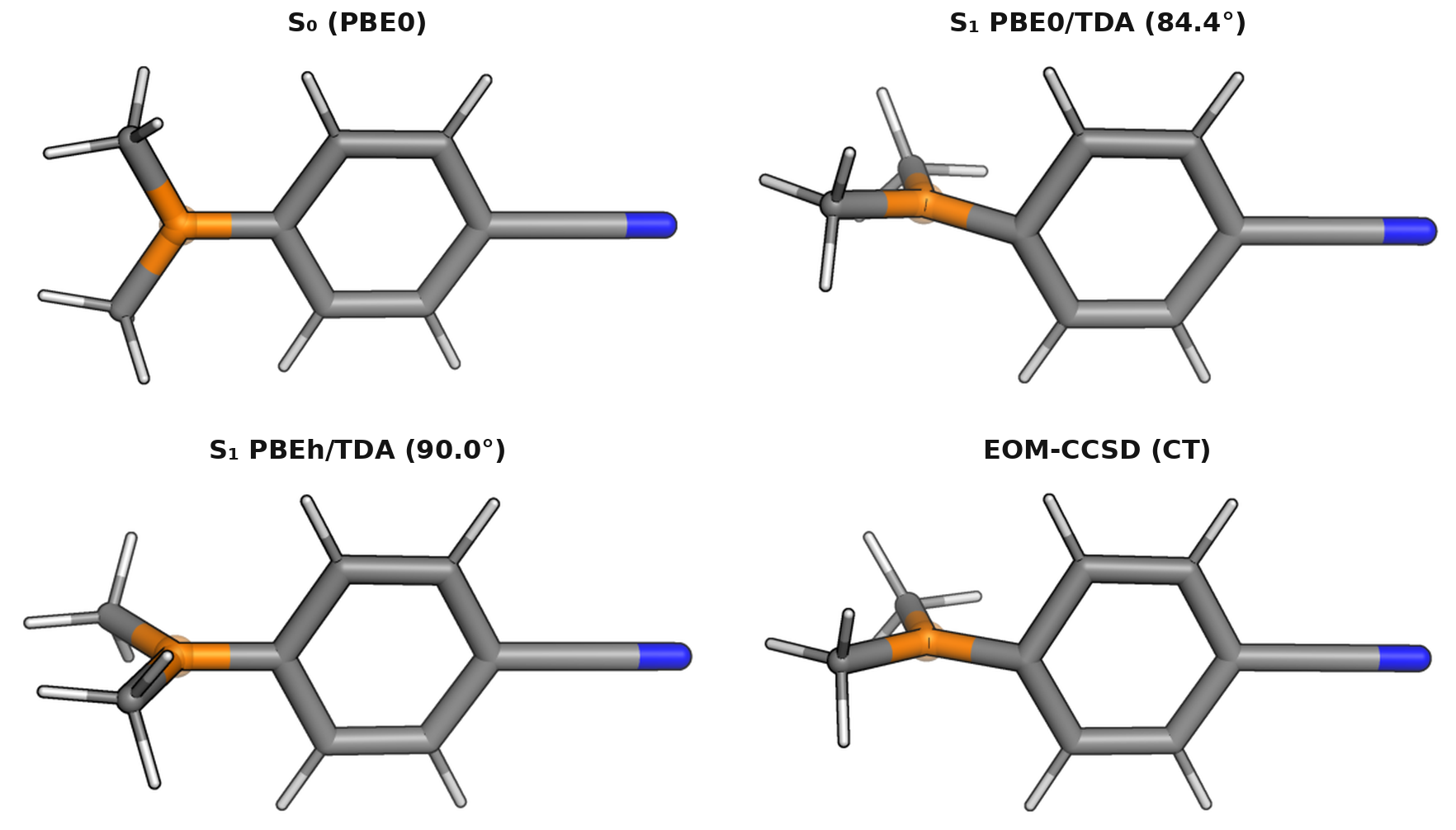}
\caption{Ground- and excited-state geometries of DMABN. Top left: PBE0/def2-SVP
ground state. Top right: $GW$-BSE@PBE0(TDA) S$_1$ minimum (twist \SI{84.4}{\degree}).
Bottom left: $GW$-BSE@PBEh(TDA) S$_1$ minimum (twist \SI{90.0}{\degree}). Bottom
right: EOM-CCSD CT reference geometry from the Supporting Information (p.~67) of
Ref.~\cite{knysh2022}. The orange sphere marks the amine nitrogen; the dimethylamino
group is planar in S$_0$ and twisted in S$_1$.}
\label{fig:dmabn_3d}
\end{figure}

\begin{figure}[t]
\centering
\includegraphics[width=0.558\linewidth]{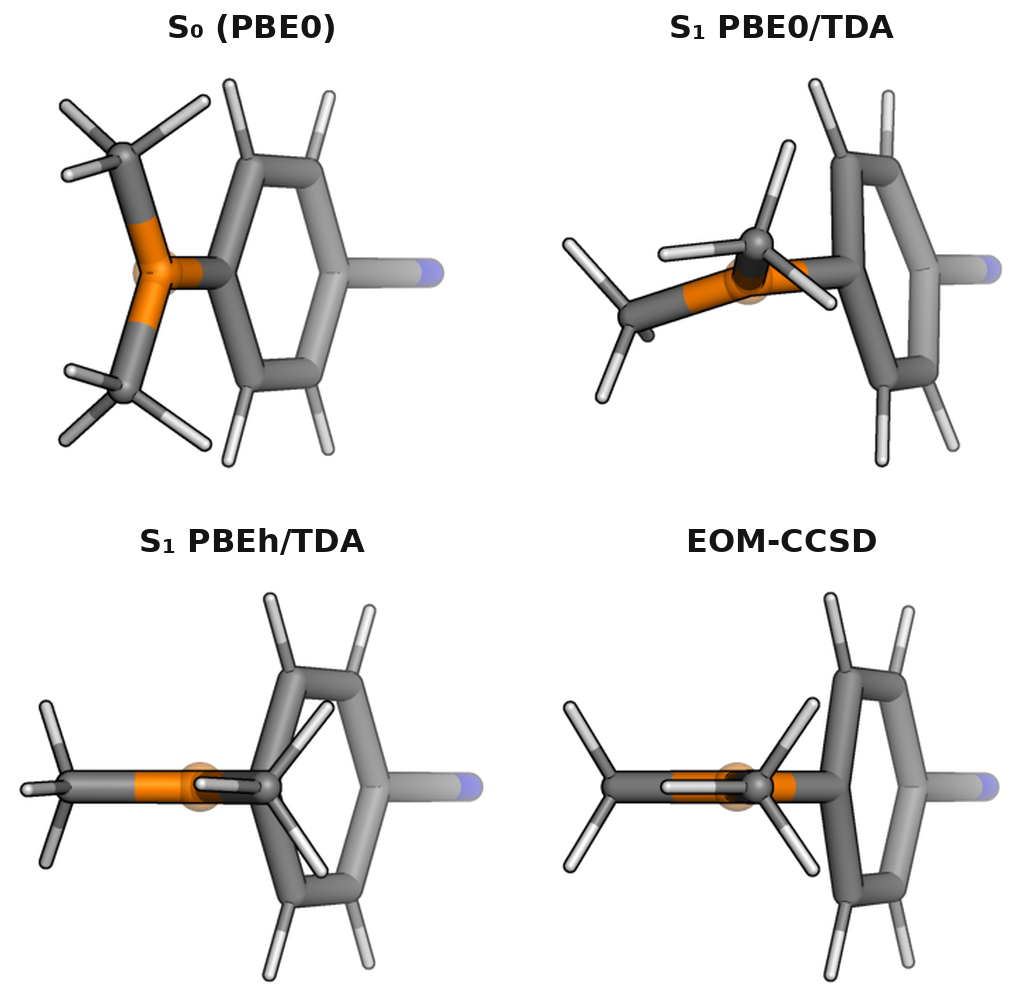}
\caption{Side views of the DMABN geometries of Fig.~\ref{fig:dmabn_3d}, obtained by
rotating the viewing direction towards the ring plane. The four structures are the
PBE0 ground state, the $GW$-BSE@PBE0(TDA) and $GW$-BSE@PBEh(TDA) S$_1$ minima, and the
EOM-CCSD CT reference geometry from the Supporting Information (p.~67) of
Ref.~\cite{knysh2022}.}
\label{fig:dmabn_3d_side}
\end{figure}

These are the structures expected from the accepted picture of DMABN: EOM-CCSD places
the charge-transfer minimum at a highly twisted geometry and finds no charge-transfer
minimum on the $\mathrm{S}_1$ surface at intermediate twists \cite{datar2023}, while
$GW$-BSE potential-energy surfaces along the same coordinate reproduce the correct
topology already for the vertical excitation energies \cite{knysh2022}. The optimized structures are shown in
Fig.~\ref{fig:dmabn_3d}, together with the EOM-CCSD CT reference geometry from the
Supporting Information (p.~67) of Ref.~\cite{knysh2022}
[J.~Chem.~Phys.\ \textbf{157}, 194102 (2022)]; side views are given in
Fig.~\ref{fig:dmabn_3d_side}. The PBEh minimum is essentially the perpendicular structure
of the wavefunction calculations, and the PBE0 one is \SI{5.6}{\degree} away from it. The qualitative behaviour of this TICT excited-state minimum is reproduced by $GW$-BSE using both starting points, and the PBEh result is in excellent quantitative agreement with the EOM-CCSD reference.

\begin{table*}[h]
\centering
\tabfont
\caption{Geometrical parameters of the DMABN ground state and of the charge-transfer
minima optimized with $GW$-BSE@PBE0(TDA), $GW$-BSE@PBEh(TDA) and EOM-CCSD. Bond
lengths in \AA, angles in degrees.}
\label{tab:dmabn}
\begin{tabular}{lcccc}
\toprule
Quantity & S$_0$ & S$_1$, PBE0 & S$_1$, PBEh & EOM-CCSD$^{c}$ \\
\midrule
twist $\theta^{a}$ / $^\circ$              & $0.5$   & $84.4$ & $90.0$ & $90.0$ \\
$R$(C$_{\rm ipso}$--N) / \AA               & $1.369$ & $1.422$ & $1.387$ & $1.431$ \\
ring alternation$^{b}$ / \AA               & $0.032$ & $0.084$ & $0.072$ & $0.077$ \\
\bottomrule
\end{tabular}
\\[3pt]
{\scriptsize $^{a}$Angle between the mean plane of the six ring carbons and the plane
of the NMe$_2$ group. $^{b}$Difference between the longest and the shortest C--C bond
of the ring. $^{c}$EOM-CCSD CT geometry from the Supporting Information (p.~67) of
Ref.~\cite{knysh2022}.}
\end{table*}

\subsection{\raggedright\normalsize Nonadiabatic dynamics of photoexcited DMABN}

The analytic gradients also make it possible to run surface-hopping trajectories of
photoexcited molecules, the application for which the present implementation of the
$GW$-BSE gradient was designed. To show that our implementation is conveniently available for the interested reader to perform such tasks, we address a physical setup studied in the TD-DFT nonadiabatic dynamics study of Mausenberger \emph{et al.}
\cite{mausenberger2025}: an ensemble of photoexcited DMABN molecules started in S$_2$,
analysed through the state populations and the ring--NMe$_2$ twist. The purpose of this
example is not to produce scientifically rigorous results, since obtaining such results requires much effort and expertise beyond the scope of this paper, but to show that our $GW$-BSE gradient can be readily plugged into a dedicated nonadiabatic-dynamics engine and used to
propagate an ensemble of molecules on the $GW$-BSE surfaces at a practical cost. The interface layer between REST and SHARC follows the design of
Ref.~\cite{mausenberger2025}: SHARC keeps the control of the nuclear and electronic
propagation and calls REST at every time step for the excitation energies, the analytic
gradients and the wave-function overlaps.

\begin{figure}[h]
\centering
\includegraphics[width=0.95\linewidth]{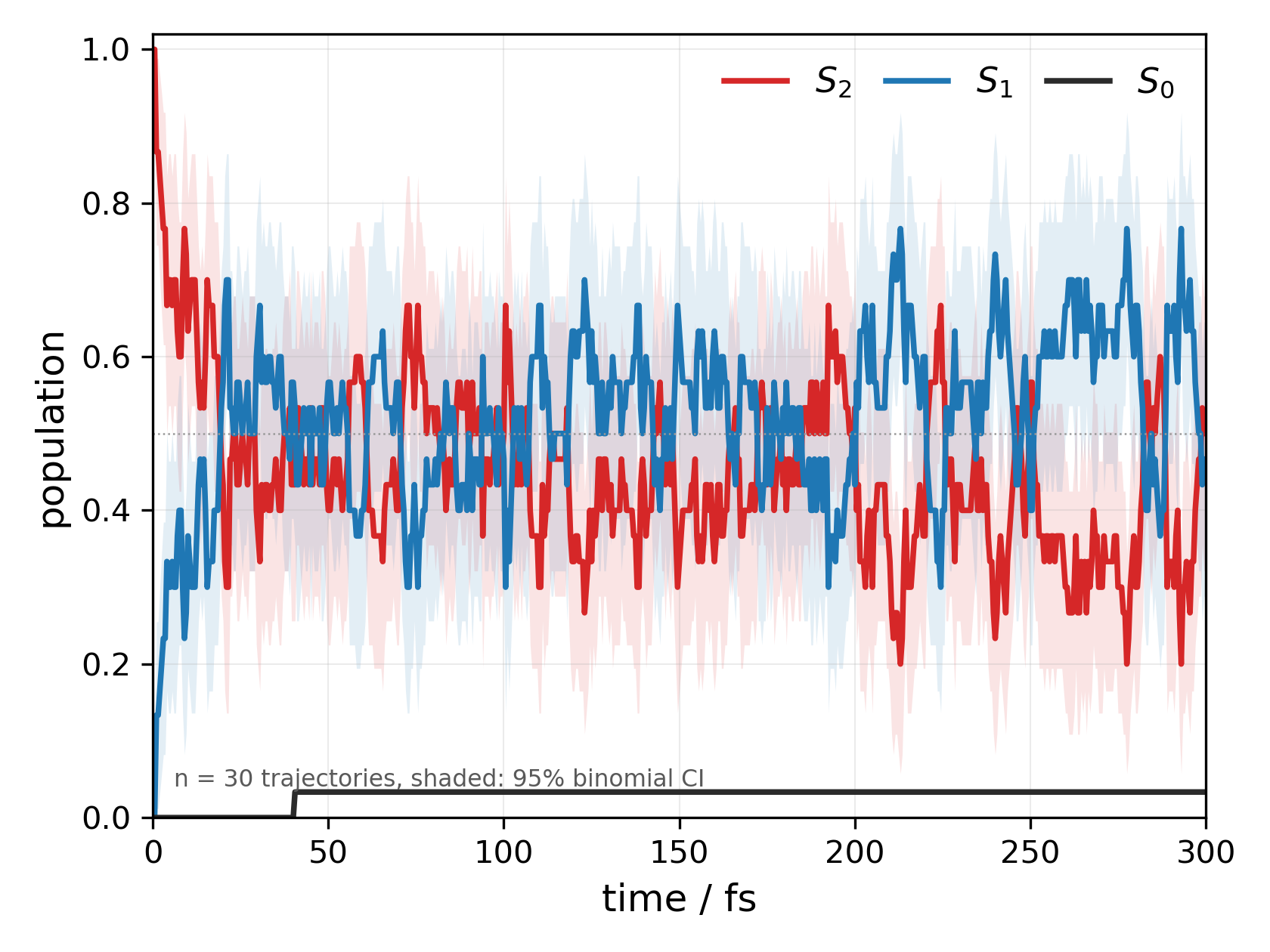}
\caption{State populations of the DMABN ensemble (S$_0$ black, S$_1$ blue, S$_2$ red),
accumulated over the \num{30} trajectories that completed the \SI{300}{fs} propagation.
Shading: 95\,\% binomial confidence interval (drawn for S$_1$ and S$_2$ only, since
$P(\mathrm{S}_0)$ is either zero or one over most of the window).}
\label{fig:dmabnpop}
\end{figure}

Fig.~\ref{fig:dmabnpop} shows the resulting populations, accumulated over the
\num{30} trajectories that have completed the propagation. The S$_2$ population decays
from unity to $0.47\pm0.20$ within \SI{25}{fs} and the S$_1$ population rises
correspondingly. After that the ensemble does not relax monotonically:
S$_1$ and S$_2$ keep exchanging population (S$_1=0.37$--$0.67$) up to \SI{300}{fs}, with
a time average over \SIrange{100}{300}{fs} of $P(\mathrm{S}_1)=0.55\pm0.09$ and
$P(\mathrm{S}_2)=0.42\pm0.09$. A single trajectory (\SI{3}{\percent}) reaches the ground
state at about \SI{40}{fs}, after which $P(\mathrm{S}_0)$ remains constant, so the
ground-state channel is open but very slow on this time scale. The repeated population transfer
reflects the near-degeneracy of the two states along the twist coordinate, which allows
the trajectories to traverse the same conical intersection region several times, which is also reported for previous NAMD studies using TD-DFT\cite{mausenberger2025}.

\begin{figure}[h]
\centering
\includegraphics[width=0.95\linewidth]{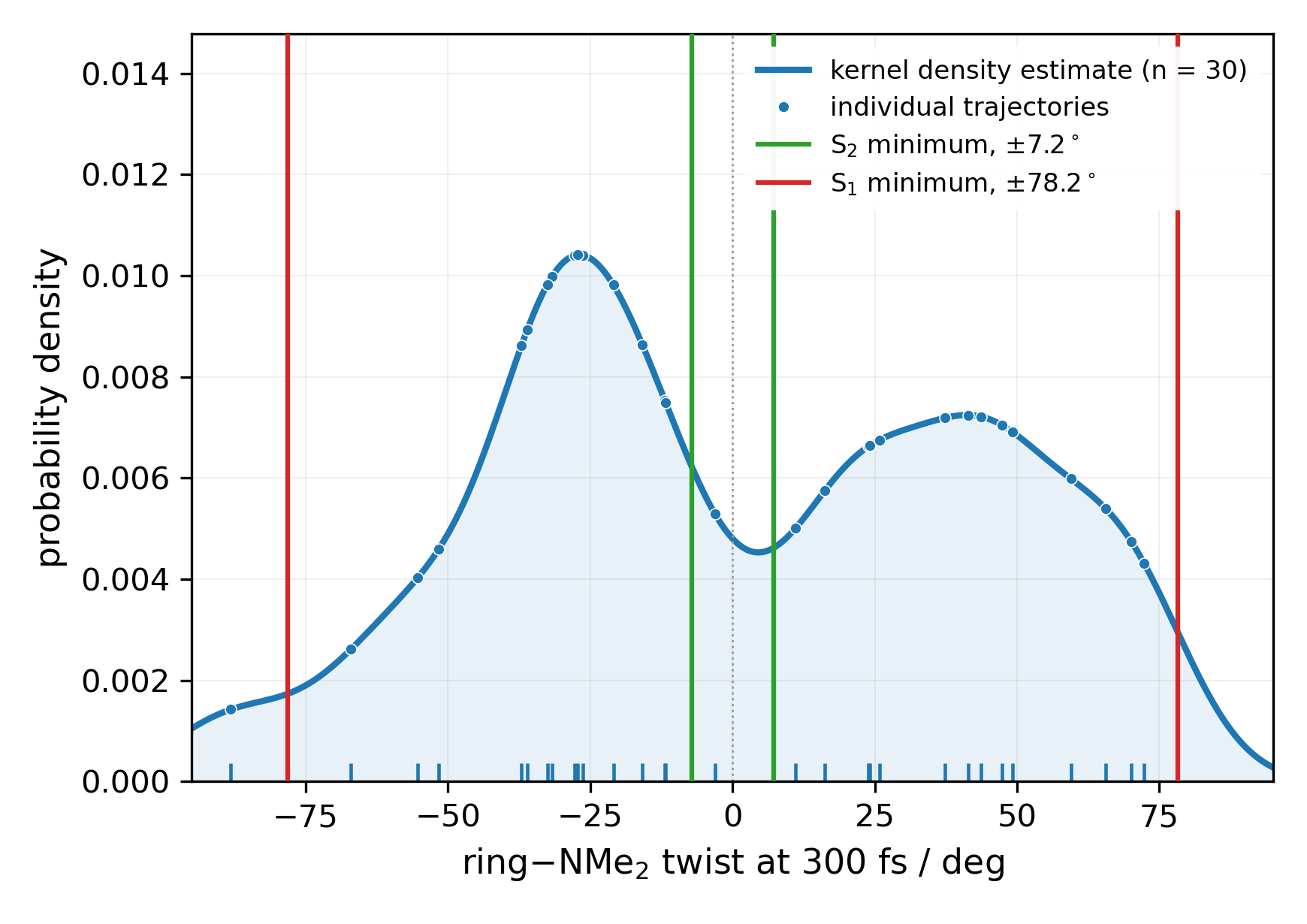}
\caption{Distribution of the ring--NMe$_2$ twist at \SI{300}{fs} over the \num{30}
trajectories that completed the propagation. The curve is a Gaussian kernel density
estimate of the \num{30} values, shown individually by the circles (on the curve) and by
the ticks along the bottom axis. The vertical lines mark the optimized S$_2$ (green)
and S$_1$ (red) minima used to interpret the distribution.}
\label{fig:dmabntwist}
\end{figure}

To characterize the structural evolution of the ensemble, we analyzed the twist angles
of all trajectories at \SI{300}{fs} and plot their distribution in
Fig.~\ref{fig:dmabntwist}. The mean absolute twist is \SI{37.7}{\degree} (median
\SI{34.3}{\degree}), only one of the \num{30} trajectories remains within
\SI{5}{\degree} of the planar structure, and \num{17} lie beyond \SI{30}{\degree}; the
distribution is symmetric with respect to the sign of the twist. To interpret this
distribution, we also optimized the S$_2$ excited-state geometry of DMABN; the S$_2$
minimum remains close to planar, while the S$_1$ minimum is strongly twisted
(Fig.~\ref{fig:dmabntwist} and Table~\ref{tab:dmabn}). Most of the trajectories lie
between these two minima, with the highest probability density between approximately
\SI{25}{\degree} and \SI{50}{\degree} in $|\theta|$. This shows that the ensemble does
not relax toward the minimum of either excited state. Instead, the molecules hop
repeatedly between S$_1$ and S$_2$ (Fig.~\ref{fig:dmabnpop}), so that the potential
energy surface and the force direction keep changing on a few-tens-of-femtoseconds time
scale, consistent with the physical picture of TSH.

\subsection{\raggedright\normalsize Performance of the implementation of the $GW$-BSE gradients}

\begin{figure*}[ht]
\centering
\includegraphics[width=0.648\textwidth]{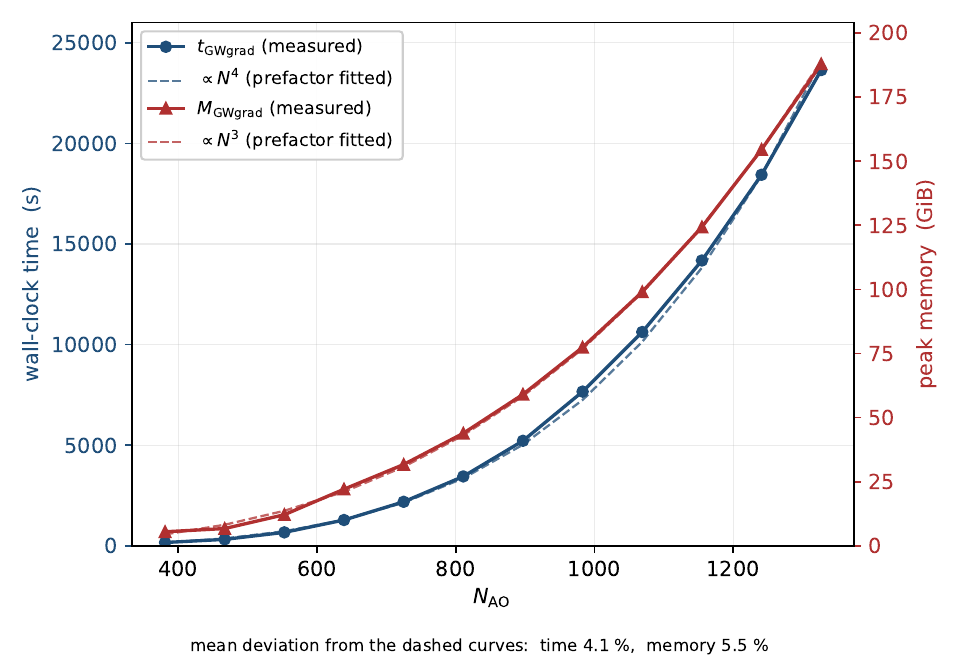}
\caption{Cost of the analytic $GW$ gradient as a function of the basis size for the
1-fluoroalkane ladder of Table~\ref{tab:perf} (AC-$GW$, no quasiparticle
extrapolation, 16 threads, one node). Left axis and filled circles: wall-clock time of
the $GW$ gradient. Right axis and filled triangles: peak memory of the $GW$ gradient.
The dashed curves' exponents are fixed to four (time) and three (memory) and only the
prefactor is adjusted by least squares in log space, so that the measured curves can
be compared directly with those powers; the mean deviation is \SI{4.1}{\percent} for
the time and \SI{5.5}{\percent} for the memory (produced by \code{plot\_scaling.py},
which reads \code{data/perf\_gw.txt}).}
\label{fig:scaling}
\end{figure*}

Table~\ref{tab:perf} reports the wall-clock time and peak memory of the two steps that
dominate the cost of a GW-BSE gradient: the $GW$ pass that produces the quasiparticle
energies, and the analytic $GW$ gradient. The ladder is the 1-fluoroalkane series
C$_m$H$_{2m+1}$F with the AC variant, on 16 threads of a dual-socket
node with two Intel Xeon E5-2673~v4 (Broadwell) processors of 20 cores each, running at
\SI{2.30}{GHz} base frequency, with \SI{440}{\giga\byte} of memory. From $m=8$ ($N_{\rm AO}=381$) to $m=30$ ($N_{\rm AO}=1327$) the $GW$
gradient grows from \SI{168}{s} to \SI{23642}{s} and its peak memory from 5.5 to
188~GiB. The computational time is well described by
\begin{equation}
t_{\rm GWgrad}\simeq c\,n_{\rm aux}^{2}\,n_{\rm mo}\,n_{\rm solve},
\qquad c\simeq 5.7~\text{ns},
\end{equation}
which reproduces the timings with $R^2=0.9991$ (maximum deviation
\SI{22}{\percent} at the smallest system, where the fixed costs are not negligible). For BSE calculations where $n_{\rm solve}$ scales linearly as the system size, this will mean that the total cost of preparing $GW$ gradients of all orbitals scale as $N^4$. The memory is dominated by the step of taking gradients of the primitive quantities and follows
$M\propto N^{2.83}$, slightly lower than the third power.
Figure~\ref{fig:scaling} shows the same data as a function of $N_{\rm AO}$ against the
$N^{4}$ and $N^{3}$ reference curves.

\begin{figure*}[ht]
\centering
\includegraphics[width=0.648\textwidth]{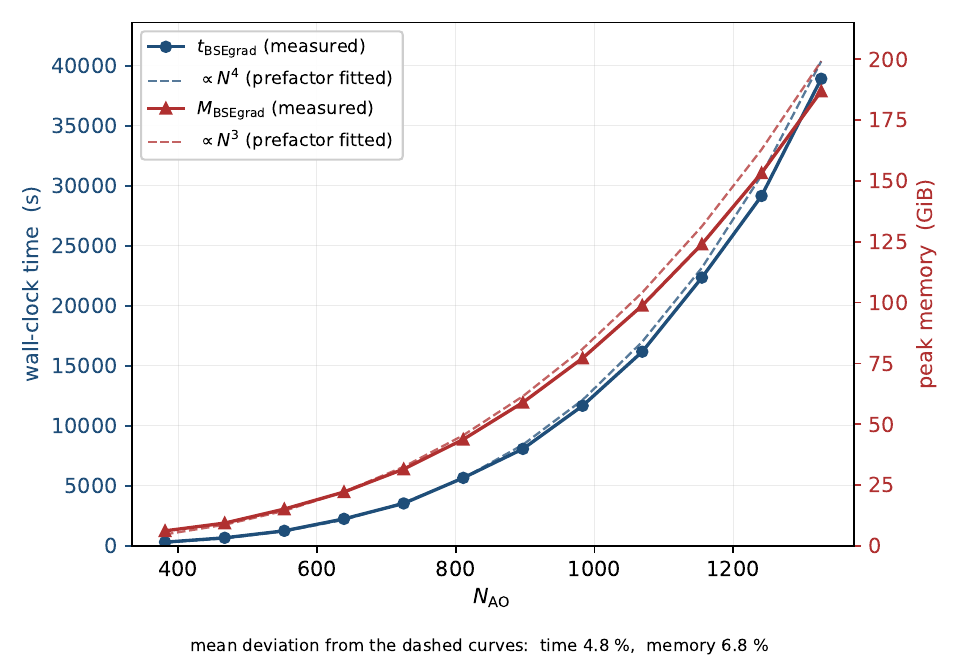}
\caption{Scaling of computational cost of the BSE gradient stage of the same
ladder (Table~\ref{tab:perf2}). This stage includes building the screening response and the BSE matvec,
solving five roots with Davidson, differentiating the lowest one and forming the
covectors; $\dim=n_{\rm occ}^{\rm act}n_{\rm vir}^{\rm act}$ is the dimension of the BSE
transition space. Filled circles (left axis): wall clock; filled triangles (right axis): peak
memory. The dashed reference curves again have their exponents fixed to four and three
and only their prefactor fitted (mean deviation \SI{4.8}{\percent} and
\SI{6.8}{\percent}); described in terms of the transition-space dimension the same data
follow $t\propto\dim^{2.0}$ and $M\propto\dim^{1.5}$ to within \SI{1.7}{\percent} and
\SI{3.8}{\percent}, consistent with the quadratic contractions that dominate the
gradient.}
\label{fig:scaling_bse}
\end{figure*}

Table~\ref{tab:perf2} reports the same quantities for the BSE gradient stage of the same
ladder, which builds the screening response and the BSE matrix, solves five roots with
Davidson, differentiates the lowest one and forms the covectors. The quantity that sets its cost is the dimension of the BSE transition space,
$\dim=n_{\rm occ}^{\rm act}n_{\rm vir}^{\rm act}$---the number of one-particle excitations retained in the $\pm0.5$~Ha window around the quasiparticle gap, which is also the size of
the matrix that the Davidson solver diagonalizes. From $m=8$
($\dim=1488$) to $m=30$ ($\dim=16728$) its wall-clock time grows from \SI{311}{s} to
\SI{38914}{s} (\SI{10.8}{h}) and its peak memory from 6.2 to 186.9~GiB. The time is well
described by
\begin{equation}
t_{\rm BSEgrad}\simeq c\,(\dim)^{2},
\qquad c\simeq 1.4\times10^{-4}~\text{s},
\end{equation}
with $R^2=0.9998$ (maximum deviation \SI{4.1}{\percent} at $m=18$).  The memory follows $M\propto(\dim)^{1.44}$, i.e.\
$M\propto N_{\rm AO}^{2.8}$, slightly below the third power and is very close to
the peak memory of the $GW$ gradient stage, because the bottleneck also lies in taking gradients of the primitive quantities. Since the BSE gradient takes
between 1.5 and 2.1 times as long as the $GW$ gradient for the same system, the two
steps together define the cost of a single gradient evaluation; the ratio decreases
smoothly with system size. Figure~\ref{fig:scaling_bse} shows the same data as a
function of $N_{\rm AO}$ against the $N^{4}$ and $N^{3}$ reference curves.

\begin{table*}[h]
\centering
\tabfont
\caption{Wall-clock times and peak memory of the $GW$ pass and of the analytic $GW$
gradient for the 1-fluoroalkane chain C$_m$H$_{2m+1}$F. $N_{\rm aux}$ is the RI auxiliary dimension,
$n_{\rm solve}$ the number of orbitals whose QP equation is solved explicitly.
$M$ is the process maximum resident set size.}
\label{tab:perf}
\begin{tabular}{lrrrrrrr}
\toprule
$m$ & $N_{\rm AO}$ & $N_{\rm aux}$ & $n_{\rm solve}$
    & $t_{\rm GW}$ (s) & $t_{\rm GWgrad}$ (s) & $M_{\rm GWgrad}$ (GiB) \\
\midrule
 8 &  381 &  983 &  86 &    90 &    168 &   5.5 \\
10 &  467 & 1205 & 104 &   133 &    323 &   6.8 \\
12 &  553 & 1427 & 122 &   234 &    671 &  12.2 \\
14 &  639 & 1649 & 141 &   403 &   1285 &  22.1 \\
16 &  725 & 1871 & 159 &   742 &   2190 &  31.7 \\
18 &  811 & 2093 & 176 &   948 &   3451 &  43.9 \\
20 &  897 & 2315 & 195 &  1525 &   5221 &  59.1 \\
22 &  983 & 2537 & 213 &  2086 &   7666 &  77.4 \\
24 & 1069 & 2759 & 231 &  3397 &  10626 &  99.0 \\
26 & 1155 & 2981 & 250 &  3596 &  14180 & 124.3 \\
28 & 1241 & 3203 & 269 &  4902 &  18439 & 154.5 \\
30 & 1327 & 3425 & 286 &  7636 &  23642 & 187.9 \\
\bottomrule
\end{tabular}
\end{table*}

\begin{table*}[h]
\centering
\tabfont
\caption{Wall-clock time, peak memory and dimension of the BSE transition space,
$\dim=n_{\rm occ}^{\rm act}n_{\rm vir}^{\rm act}$ (pairs inside a $\pm0.5$~Ha window around
the quasiparticle gap), for the BSE gradient of the 1-fluoroalkane ladder (def2-TZVP, PBEh, 16 threads, one job at a
time, $\pm0.5$~Ha transition window). This timing includes the differentiation of the lowest excited state and the
evaluation of the covectors; its cost is dominated by the contractions over the
transition space.}
\label{tab:perf2}
\begin{tabular}{lrrrrr}
\toprule
$m$ & $N_{\rm AO}$ & $\dim$ & $t_{\rm BSEgrad}$ (s) & $M_{\rm BSEgrad}$ (GiB) \\
\midrule
 8 &  381 &  1488 &    311 &   6.2 \\
10 &  467 &  2220 &    673 &   9.4 \\
12 &  553 &  3045 &   1254 &  15.1 \\
14 &  639 &  4040 &   2252 &  22.2 \\
16 &  725 &  5130 &   3535 &  31.6 \\
18 &  811 &  6300 &   5667 &  43.8 \\
20 &  897 &  7784 &   8081 &  59.0 \\
22 &  983 &  9272 &  11672 &  77.2 \\
24 & 1069 & 10890 &  16178 &  98.8 \\
26 & 1155 & 12709 &  22351 & 124.0 \\
28 & 1241 & 14784 &  29153 & 153.3 \\
30 & 1327 & 16728 &  38914 & 186.9 \\
\bottomrule
\end{tabular}
\end{table*}

\section{\raggedright\large Conclusion}

We have derived, implemented and validated analytic nuclear gradients for three variants of
$GW$ self-energies and for the Bethe-Salpeter excitation energies built on top of them. The
gradients are formulated on a small set of primitives---orbital energies, the RI
three-centre tensor, the Coulomb metric, and the diagonal exchange-correlation
potential. Covectors of the primitive derivatives are derived through a back-propagation procedure and efficiently assembled into the final gradient. The implementation is available in the open-source code.

The implementation was verified against finite differences: the three $GW$ gradients
agree with central differences to \num{2e-7}--\num{6e-5}~Ha/Bohr, they agree with each
other to \num{1e-7}--\num{5e-4}~Ha/Bohr. With the AC-$GW$-BSE gradient we relaxed excited states of CO, a donor-acceptor dye DA1, and the well-studied system of DMABN, reproducing the structural
changes observed by experiments or by high-level wavefunction methods. We measured the cost of the evaluating gradients of AC-$GW$ and BSE, confirming that the computational time scales no greater than $N^4$ and memory less than $N^3$. Together with
the Supporting Information, which derives the CD and CD-WAC channels, these results make
excited-state geometry optimization and
minimum-energy-path searches at the $GW$-BSE level a routine tool for molecules of
moderate size.
Interfacing REST with SHARC, we further
propagated an ensemble of photoexcited DMABN molecules on the AC-$GW$-BSE surfaces and
reproduced the two fingerprints of its charge-transfer photochemistry: an essentially
barrierless S$_2\to$S$_1$ transfer on a \SI{10}{fs} time scale and the depopulation of the
planar structure in favour of twisted geometries. 

For future work, we plan to extend the availability of analytic gradients of $GW$ to ev$GW$, and further benchmark the optmized geometries from different levels of $GW$-BSE against experimental and high-level quantum chemistry data. Since our derivation of $GW$-BSE employs a clean and fundamental back-propagation methodology, we will also try employing the same methodology to deriving analytic gradients of more quantum chemical methods. We will also explore the use of $GW$-BSE in NAMD simulations of larger systems, and investigate in further improving the performance of the implementation on high-performance computing platforms.

\section*{Acknowledgment}
We are grateful for support from the National Natural Science Foundation of China (Grant Nos. 22125301, 22393912, 22321003), Quantum Science and Technology---National Science and Technology Major Project (Project No. 2021ZD0303305), and the Robotic AI-Scientist platform of the Chinese Academy of Sciences.


{\footnotesize\begin{thebibliography}{99}

\bibitem{pulay1969} P.~Pulay, \emph{Ab initio calculation of force constants and
equilibrium geometries in polyatomic molecules}, Mol.\ Phys.\ \textbf{17}, 197 (1969).
\doi{10.1080/00268976900100941}

\bibitem{schlegel1982} H.~B.~Schlegel, \emph{Optimization of equilibrium geometries and
transition structures}, J.\ Comput.\ Chem.\ \textbf{3}, 214 (1982).
\doi{10.1002/jcc.540030212}

\bibitem{bernardi1996} F.~Bernardi, M.~Olivucci, M.~A.~Robb, \emph{Potential energy
surface crossings in organic photochemistry}, Chem.\ Soc.\ Rev.\ \textbf{25}, 321 (1996).
\doi{10.1039/CS9962500321}

\bibitem{vancaillie1999} C.~Van Caillie, R.~D.~Amos, \emph{Geometric derivatives of
excitation energies using SCF and DFT}, Chem.\ Phys.\ Lett.\ \textbf{308}, 249 (1999).
\doi{10.1016/S0009-2614(99)00646-6}

\bibitem{furche2002} F.~Furche, R.~Ahlrichs, \emph{Adiabatic time-dependent density
functional methods for excited state properties}, J.\ Chem.\ Phys.\ \textbf{117}, 7433
(2002). \doi{10.1063/1.1508368}

\bibitem{scalmani2006} G.~Scalmani, M.~J.~Frisch, B.~Mennucci, J.~Tomasi, R.~Cammi,
V.~Barone, \emph{Geometries and properties of excited states in the gas phase and in
solution}, J.\ Chem.\ Phys.\ \textbf{124}, 094107 (2006). \doi{10.1063/1.2173258}

\bibitem{send2010} R.~Send, F.~Furche, \emph{First-order nonadiabatic couplings from
time-dependent hybrid density functional response theory}, J.\ Chem.\ Phys.\ \textbf{132},
044107 (2010). \doi{10.1063/1.3292571}

\bibitem{cembran2004} C.~Cembran, F.~Bernardi, M.~Garavelli, L.~Gagliardi, G.~Orlandi,
\emph{On the mechanism of the cis-trans isomerization in the lowest electronic states of
azobenzene: S$_0$, S$_1$, and T$_1$}, J.\ Am.\ Chem.\ Soc.\ \textbf{126}, 3234 (2004).
\doi{10.1021/ja038327y}

\bibitem{tapavicza2007} E.~Tapavicza, I.~Tavernelli, U.~R\"othlisberger, \emph{Trajectory
surface hopping within linear response time-dependent density-functional theory},
Phys.\ Rev.\ Lett.\ \textbf{98}, 023001 (2007). \doi{10.1103/PhysRevLett.98.023001}

\bibitem{dreuw2005} A.~Dreuw, M.~Head-Gordon, \emph{Single-reference ab initio methods for
the calculation of excited states of large molecules}, Chem.\ Rev.\ \textbf{105}, 4009
(2005). \doi{10.1021/cr0505627}

\bibitem{furche2005} F.~Furche, D.~Rappoport, \emph{Density functional methods for excited
states: equilibrium structure and electronic spectra}, in \emph{Computational
Photochemistry} (Theoretical and Computational Chemistry, Vol.~16), Elsevier, 2005.
\doi{10.1016/S1380-7323(05)80020-2}

\bibitem{onida2002} G.~Onida, L.~Reining, A.~Rubio, \emph{Electronic excitations:
density-functional versus many-body Green's-function approaches}, Rev.\ Mod.\ Phys.\
\textbf{74}, 601 (2002). \doi{10.1103/RevModPhys.74.601}

\bibitem{rohlfing2000} M.~Rohlfing, S.~G.~Louie, \emph{Electron-hole excitations and
optical spectra from first principles}, Phys.\ Rev.\ B \textbf{62}, 4927 (2000).
\doi{10.1103/PhysRevB.62.4927}

\bibitem{blase2018} X.~Blase, I.~Duchemin, D.~Jacquemin, \emph{The Bethe--Salpeter equation
in chemistry: relations with TD-DFT, applications and challenges}, Chem.\ Soc.\ Rev.\
\textbf{47}, 1022 (2018). \doi{10.1039/C7CS00049A}

\bibitem{jacquemin2015} D.~Jacquemin, I.~Duchemin, X.~Blase, \emph{0-0 energies using
hybrid schemes: benchmarks of TD-DFT, CIS(D), ADC(2), CC2 and BSE/GW formalisms for 80
real-life compounds}, J.\ Chem.\ Theory Comput.\ \textbf{11}, 5340 (2015).
\doi{10.1021/acs.jctc.5b00619}

\bibitem{ismail2003} S.~Ismail-Beigi, S.~G.~Louie, \emph{Excited-state forces within a
first-principles Green's function formalism}, Phys.\ Rev.\ Lett.\ \textbf{90}, 076401
(2003). \doi{10.1103/PhysRevLett.90.076401}

\bibitem{villalobos2023} J.~Villalobos-Castro, I.~Knysh, D.~Jacquemin, I.~Duchemin,
X.~Blase, \emph{Lagrangian Z-vector approach to Bethe--Salpeter analytic gradients:
assessing approximations}, J.\ Chem.\ Phys.\ \textbf{159}, 024116 (2023).
\doi{10.1063/5.0156687}

\bibitem{caylak2021} \c{C}aylak, D.~Baumeier, \emph{Excited-state geometry optimization of
small molecules with many-body Green's function theory}, J.\ Chem.\ Theory Comput.\
\textbf{17}, 879 (2021). \doi{10.1021/acs.jctc.0c01099}

\bibitem{alrahamneh2025} M.~Alrahamneh, I.~Tolbatov, P.~Umari, \emph{Excited-state forces
with $GW$-BSE through the Hellmann--Feynman theorem}, Int.\ J.\ Mol.\ Sci.\ \textbf{26},
2306 (2025). \doi{10.3390/ijms26052306}

\bibitem{delgrande2025} R.~R.~Del Grande, D.~A.~Strubbe, \emph{Revisiting ab-initio
excited state forces from many-body Green's function formalism: approximations and
benchmark}, arXiv:2502.05144 (2025). \doi{10.48550/arXiv.2502.05144}

\bibitem{nist_co_diatomic} K.~P.~Huber, G.~H.~Herzberg, \emph{Constants of Diatomic
Molecules}, in \emph{NIST Chemistry WebBook, NIST Standard Reference Database Number
69}, edited by P.~J.~Linstrom and W.~G.~Mallard (National Institute of Standards and
Technology, Gaithersburg, MD). \doi{10.18434/T4D303}. Data prepared by J.~W.~Gallagher
and R.~D.~Johnson III; carbon monoxide entry, accessed November 13, 2026.

\bibitem{tolleG0W0} J.~T\"olle, \emph{Fully analytic $G_0W_0$ nuclear gradients},
J.\ Phys.\ Chem.\ Lett.\ \textbf{16}, 3672 (2025). \doi{10.1021/acs.jpclett.5c00046}

\bibitem{tolle2025} J.~T\"olle, M.-P.~Kitsaras, P.-F.~Loos, \emph{Fully analytic nuclear
gradients for the Bethe--Salpeter equation}, J.\ Phys.\ Chem.\ Lett.\ \textbf{16}, 11134
(2025). \doi{10.1021/acs.jpclett.5c02219}

\bibitem{holzer2026} C.~Holzer, \emph{Analytic $GW$-Bethe--Salpeter gradients with
quasiparticle response in the resolution of the identity}, arXiv:2611.12617 (2026).
\doi{10.48550/arXiv.2611.12617}

\bibitem{faber2014} C.~Faber, P.~Boulanger, C.~Attaccalite, I.~Duchemin, X.~Blase,
\emph{Excited states properties of organic molecules: from density functional theory to
the GW and Bethe--Salpeter Green's function formalisms}, Phil.\ Trans.\ R.\ Soc.\ A
\textbf{372}, 20130271 (2014). \doi{10.1098/rsta.2013.0271}

\bibitem{knysh2022} I.~Knysh, I.~Duchemin, X.~Blase, D.~Jacquemin, \emph{Modeling of
excited state potential energy surfaces with the Bethe--Salpeter equation formalism: the
4-(dimethylamino)benzonitrile twist}, J.\ Chem.\ Phys.\ \textbf{157}, 194102 (2022).
\doi{10.1063/5.0121121}

\bibitem{knysh2023} I.~Knysh, K.~Letellier, I.~Duchemin, X.~Blase, D.~Jacquemin,
\emph{Excited state potential energy surfaces of N-phenylpyrrole upon twisting: reference
values and comparison between BSE/GW and TD-DFT}, Phys.\ Chem.\ Chem.\ Phys.\ \textbf{25},
8376 (2023). \doi{10.1039/D3CP00474K}

\bibitem{rest2025} Z.~Li, T.~Gao, S.~Wang, S.~Bi, R.~Feng, Z.~Zhu, Y.~Zhao, W.~Yan, L.~Yu,
Q.~Gao, Z.~Lin, J.~Wu, I.~Y.~Zhang, X.~Xu, \emph{REST: embracing the Rust programming
language for modern electronic structure theory}, Chin.\ J.\ Chem.\ Phys.\ \textbf{38},
788 (2025). \doi{10.1063/1674-0068/cjcp2510156}

\bibitem{restgwbse2026} Q.~Gao, I.~Y.~Zhang, X.~Xu, \emph{GW-BSE for molecular excited
states in REST: state-of-the-art methods, acceleration strategies, and the LAMB
approximation}, J.\ Chem.\ Phys.\ \textbf{164}, 214116 (2026).
\doi{10.1063/5.0331920}

\bibitem{hartree1928} D.~R.~Hartree, \emph{The wave mechanics of an atom with a non-Coulomb
central field. Part I}, Math.\ Proc.\ Cambridge Philos.\ Soc.\ \textbf{24}, 89 (1928).
\doi{10.1017/S0305004100011919}

\bibitem{fock1930} V.~Fock, \emph{N\"aherungsmethode zur L\"osung des quantenmechanischen
Mehrk\"orperproblems}, Z.\ Phys.\ \textbf{61}, 126 (1930). \doi{10.1007/BF01340294}

\bibitem{roothaan1951} C.~C.~J.~Roothaan, \emph{New developments in molecular orbital
theory}, Rev.\ Mod.\ Phys.\ \textbf{23}, 69 (1951). \doi{10.1103/RevModPhys.23.69}

\bibitem{perdew1996} J.~P.~Perdew, K.~Burke, M.~Ernzerhof, \emph{Generalized gradient
approximation made simple}, Phys.\ Rev.\ Lett.\ \textbf{77}, 3865 (1996).
\doi{10.1103/PhysRevLett.77.3865}

\bibitem{adamo1999} C.~Adamo, V.~Barone, \emph{Toward reliable density functional methods
without adjustable parameters: the PBE0 model}, J.\ Chem.\ Phys.\ \textbf{110}, 6158
(1999). \doi{10.1063/1.478522}

\bibitem{weigend2005} F.~Weigend, R.~Ahlrichs, \emph{Balanced basis sets of split valence,
triple zeta valence and quadruple zeta valence quality for H to Rn}, Phys.\ Chem.\ Chem.\
Phys.\ \textbf{7}, 3297 (2005). \doi{10.1039/b508541a}

\bibitem{weigend2006} F.~Weigend, \emph{Accurate Coulomb-fitting basis sets for H to Rn},
Phys.\ Chem.\ Chem.\ Phys.\ \textbf{8}, 1057 (2006). \doi{10.1039/b515623h}

\bibitem{vahtras1993} O.~Vahtras, J.~Alml\"of, M.~W.~Feyereisen, \emph{Integral
approximations for LCAO-SCF calculations}, Chem.\ Phys.\ Lett.\ \textbf{213}, 514 (1993).
\doi{10.1016/0009-2614(93)89151-7}

\bibitem{eichkorn1995} K.~Eichkorn, O.~Treutler, H.~\"Ohm, M.~H\"aser, R.~Ahlrichs,
\emph{Auxiliary basis sets to approximate Coulomb potentials}, Chem.\ Phys.\ Lett.\
\textbf{240}, 283 (1995). \doi{10.1016/0009-2614(95)00621-A}

\bibitem{bruneval2013} F.~Bruneval, M.~A.~L.~Marques, \emph{Benchmarking the starting
points of the GW approximation for molecules}, J.\ Chem.\ Theory Comput.\ \textbf{9}, 324
(2013). \doi{10.1021/ct300835h}

\bibitem{vansetten2015} M.~J.~van Setten, F.~Caruso, S.~Sharifzadeh, X.~Ren, M.~Scheffler,
F.~Liu, J.~Lischner, L.~Lin, J.~R.~Deslippe, S.~G.~Louie, C.~Yang, F.~Weigend,
J.~B.~Neaton, F.~Evers, P.~Rinke, \emph{GW100: benchmarking $G_0W_0$ for molecular
systems}, J.\ Chem.\ Theory Comput.\ \textbf{11}, 5665 (2015).
\doi{10.1021/acs.jctc.5b00453}

\bibitem{bois2017} J.~Bois, T.~K\"orzd\"orfer, \emph{Size dependence of nonempirically
tuned DFT starting points for $G_0W_0$ applied to $\pi$-conjugated molecular chains},
J.\ Chem.\ Theory Comput.\ \textbf{13}, 4962 (2017). \doi{10.1021/acs.jctc.7b00557}

\bibitem{golze2018} D.~Golze, J.~Wilhelm, M.~J.~van Setten, P.~Rinke, \emph{Core-level
binding energies from $GW$: an efficient full-frequency approach within a localized
basis}, J.\ Chem.\ Theory Comput.\ \textbf{14}, 4856 (2018).
\doi{10.1021/acs.jctc.8b00458}

\bibitem{panades2023} R.~L.~Panad\'es-Barrueta, D.~Golze, \emph{Accelerating core-level
$GW$ calculations by combining the contour deformation approach with the analytic
continuation of $W$}, J.\ Chem.\ Theory Comput.\ \textbf{19}, 5450 (2023).
\doi{10.1021/acs.jctc.3c00555}

\bibitem{tapavicza2009} E.~Tapavicza, I.~Tavernelli, U.~R\"othlisberger, \emph{Ab initio
excited state properties and dynamics of a prototype $\sigma$-bridged donor--acceptor
molecule}, J.\ Phys.\ Chem.\ A \textbf{113}, 9595 (2009). \doi{10.1021/jp901356k}

\bibitem{vidberg1977} H.~J.~Vidberg, J.~W.~Serene, \emph{Solving the Eliashberg equations
by means of N-point Pade approximants}, J.\ Low Temp.\ Phys.\ \textbf{29}, 179 (1977).
\doi{10.1007/BF00655090}

\bibitem{ren2012} X.~Ren, P.~Rinke, V.~Blum, J.~Wieferink, A.~Tkatchenko, A.~Sanfilippo,
K.~Reuter, M.~Scheffler, \emph{Resolution-of-identity approach to Hartree--Fock, hybrid
density functionals, RPA, MP2 and $GW$ with numeric atom-centered orbital basis
functions}, New J.\ Phys.\ \textbf{14}, 053020 (2012). \doi{10.1088/1367-2630/14/5/053020}

\bibitem{schindlmayr1998} A.~Schindlmayr, T.~J.~Pollehn, R.~W.~Godby, \emph{Spectra and
total-energy functionals from an equation-of-motion theory}, Phys.\ Rev.\ B \textbf{58},
12684 (1998). \doi{10.1103/PhysRevB.58.12684}

\bibitem{budzak2017} \v{S}.~Budz\'ak, M.~Fumanal, \emph{et al.}, \emph{CC3/CASPT2
benchmark of excited-state geometries of small carbonyls}, J.\ Chem.\ Theory Comput.\
\textbf{13}, 6237 (2017).

\bibitem{geometric2016} L.-P.~Wang, C.~C.~Song, \emph{Geometry optimization made simple
with translation and rotation coordinates}, J.~Chem.~Phys. \textbf{144}, 214108 (2016).

\doi{10.1063/1.4952956}
\bibitem{richter2011} M.~Richter, P.~Marquetand, J.~Gonz\'alez-V\'azquez, I.~Sol\'a,
L.~Gonz\'alez, \emph{SHARC: ab initio molecular dynamics with surface hopping in the
adiabatic representation including arbitrary couplings}, J.~Chem.~Theory Comput.
\textbf{7}, 1253--1258 (2011). \doi{10.1021/ct1007394}

\bibitem{sharc2018} S.~Mai, P.~Marquetand, L.~Gonz\'alez, \emph{Nonadiabatic dynamics:
the SHARC approach}, WIREs Comput.~Mol.~Sci. \textbf{8}, e1370 (2018).

\doi{10.1002/wcms.1370}
\bibitem{sharc2026} S.~Mai, B.~Bachmair, L.~Gagliardi, \emph{et al.}, \emph{SHARC
Release 4.1}, Zenodo (2026). \doi{10.5281/zenodo.20927309}

\bibitem{mausenberger2025} S.~Mausenberger, S.~Polonius, S.~Mai, L.~Gonz\'alez,
\emph{Efficient, hierarchical, and object-oriented electronic structure interfaces for
direct nonadiabatic dynamics simulations}, J.~Chem.~Theory Comput. \textbf{21},
8994--9008 (2025).

\doi{10.1021/acs.jctc.5c00878}
\bibitem{grabowski2003} Z.~R.~Grabowski, K.~Rotkiewicz, W.~Rettig, \emph{Structural
changes accompanying intramolecular electron transfer: focus on twisted
intramolecular charge-transfer states and structures}, Chem.~Rev. \textbf{103},
3899--4032 (2003).
\doi{10.1021/cr940269m}

\bibitem{datar2023} A.~Datar, S.~Gudivada, D.~A.~Matthews, \emph{Ab initio investigation
of intramolecular charge transfer states in DMABN by calculation of excited state X-ray
absorption spectra}, J.~Phys.~Chem.~A \textbf{127}, 4643--4649 (2023).
\doi{10.1021/acs.jpca.3c01409}

\bibitem{zhang2022} X.~Zhang, G.~K.-L.~Chan, \emph{Differentiable quantum chemistry with
PySCF for molecules and materials at the mean-field level and beyond},
J.\ Chem.\ Phys.\ \textbf{157}, 204801 (2022). \doi{10.1063/5.0118200}

\bibitem{song2021} C.~Song, T.~J.~Mart\'inez, J.~B.~Neaton, \emph{A diagrammatic approach
for automatically deriving analytical gradients of tensor hyper-contracted electronic
structure methods}, J.\ Chem.\ Phys.\ \textbf{155}, 024108 (2021).
\doi{10.1063/5.0055914}

\bibitem{mitchell2024} E.~C.~Mitchell, J.~M.~Turney, H.~F.~Schaefer, \emph{Automatic
differentiation for explicitly correlated MP2}, J.\ Chem.\ Theory Comput.\ \textbf{20},
8529--8538 (2024). \doi{10.1021/acs.jctc.4c00818}

\end{thebibliography}}
\end{document}


\maketitle

\noindent The imaginary-axis self-energy with its Pad\'e continuation is the
resolution-of-identity $GW$ scheme of Ren \emph{et al.}\ [New J. Phys. \textbf{14}, 053020
(2012)]. The contour-deformation (CD)
self-energy used throughout is the full-frequency
localized-basis formulation of Golze \emph{et al.}\ [J. Chem. Theory Comput. \textbf{14},
4856 (2018)], and the CD-WAC variant of
Sec.~\ref{sec:wac} is the contour-deformation/analytic-continuation scheme of
Panad\'es-Barrueta and Golze [J. Chem. Theory Comput. \textbf{19}, 5450 (2023)], adapted here so that its gradient can be
expressed through the same covectors.

\noindent This Supporting Information collects (S1) the Pad\'e approximant in AC and CD-WAC and its Jacobian, (S2) the
contour-deformation (CD) self-energy and its residue channel, (S3) the static subtraction
that regularizes the residue crossing, (S4) the analytic gradient of the CD self-energy,
(S5) the CD-WAC variant, (S6) the analytic gradient of the CD-WAC self-energy, and (S7)
the final relaxed geometries of DA1 at DMABN at the ground state and at relavent excited-states.

\section{The Pad\'e approximant in AC and CD-WAC and its Jacobian}\label{sec:ac}

\paragraph{Sample selection.}
The analytic continuation strategy---evaluate $\Sigma^{c}(iu)$ on the imaginary axis and obtain
the real-frequency self-energy by a Pad\'e approximant---follows the
resolution-of-identity $GW$ implementation of Ren \emph{et al.}\ [New J. Phys.
\textbf{14}, 053020 (2012)].
The imaginary-axis quadrature is a modified Gauss--Legendre grid on $[0,\infty)$
(transformed to $[0,1]$) whose nodes depend only on the reference spectrum and are held
fixed under nuclear displacement. The approximant is built on the Fermi-shifted axis,
i.e.\ from the samples $S_k=\Sigma^{c}_{nn}(\xi_k)$ at
$\xi_k=e_F+i\lambda_k$ with $e_F=\tfrac12(\epsilon_{\rm HOMO}+\epsilon_{\rm LUMO})$, the
shift used by the resolution-of-identity implementation; every node moves with the Fermi
level, $\partial\xi_k/\partial e_F=1$, and the gradient carries that dependence explicitly
(below). Of the $20$ nodes of the production grid, $16$ are used as the input of the
Pad\'e fit, selected with the geometric spacing of the reference implementation
(\texttt{ac\_num\_samples}; the step decreases from $1$ to $\tfrac23$ of the mean), so that
both the small- and the large-frequency part of the grid are represented; the four unused
nodes serve as a consistency check (their reconstructed self-energy agrees with the direct
evaluation to better than $10^{-8}$~Ha in all tests reported here).

\paragraph{Continued fraction.}
With the samples $S_k=\Sigma^{c}_{nn}(\xi_k)$ at the nodes $\xi_k=e_F+i\lambda_k$, the
Pad\'e approximant is written as the continued fraction
\begin{equation}\label{eq:pade}
\mathcal{P}(z)=a_0+\frac{z-\xi_0}{a_1+\dfrac{z-\xi_1}{a_2+\cdots}},
\qquad z=\omega+i\eta_{\rm ac},
\end{equation}
whose coefficients $a_k$ are obtained from the samples by the Thiele reciprocal-difference
recursion (the same recursion as Eq.~\eqref{eq:wacjeff} below, with the nodes $\xi_k$). The
QP root is then the solution of
$\omega=\epsilon_n+\Sigma^x_{nn}-v^{xc}_n
+\mathrm{Re}\,\mathcal{P}(\omega)$, found by Newton iteration; the
renormalization factor of the main text is
\begin{equation}\label{eq:Z}
Z_n^{-1}=1-\mathrm{Re}\,\frac{\partial\mathcal{P}}{\partial\omega}\Big|_{\omega_n},
\end{equation}
where the derivative is evaluated analytically from the continued fraction (not by finite
differences of the root).

\paragraph{Jacobian of the continuation.}
Writing the approximant as a function of its samples and of the Fermi shift,
$\mathcal{P}=\mathcal{P}(z;\{S_k\},e_F)$, the fixed-$\omega$ derivative required by the
gradient is
\begin{equation}\label{eq:jac}
\frac{\partial\mathcal{P}}{\partial\xi}\Big|_{\omega}
=\sum_k J_k(\omega)\,\frac{\text{d}S_k}{\text{d}\xi}
+J_{e_F}(\omega)\,\frac{\text{d}e_F}{\text{d}\xi},
\qquad J_k(\omega)=\frac{\partial\mathcal{P}}{\partial S_k}\Big|_{\omega},
\quad J_{e_F}(\omega)=\frac{\partial\mathcal{P}}{\partial e_F}\Big|_{\omega},
\end{equation}
and $J_k$ is obtained by differentiating the recursion that determines the $a_k$. Each
$a_k$ is a smooth function of $S_0,\dots,S_k$ and of the nodes $\xi_0,\dots,\xi_k$, so
$\partial a_k/\partial S_j$ and $\partial a_k/\partial e_F$ follow by differentiating that
recursion term by term---a short chain of additions, multiplications and divisions---and
the evaluation of the continued fraction is differentiated with the same algebra; no finite
differences are involved anywhere. Because every node moves with the Fermi level, the
approximant depends on $e_F$ both through the nodes and through the sample values, so the
chain rule of Eq.~\eqref{eq:jac} carries the Fermi-level column $J_{e_F}$ in addition to
the sample sum. The resulting $n_{\rm sample}\times n_{\rm sample}$ matrix $J$ (plus its
$e_F$ column) is dense but small (16$\times$16 in the production setup) and is built once
per geometry and per target orbital. This is the entire additional cost of the AC gradient
with respect to a direct frequency-by-frequency treatment: no primitive derivative is
recomputed, because the sample derivatives $\text{d}S_k/\text{d}\xi$ are the contractions
of the correlation covectors in Eqs.~(16)--(24) of the main text with the node coefficient
$c_{nk}$ replaced by the sample kernel at $\xi_k$.

\section{The CD self-energy and its residue channel}\label{sec:cd}

\paragraph{Deformation of the frequency integral.}
The correlation self-energy of orbital $n$ is
\begin{equation}\label{eq:deform}
\Sigma^{c}_{nn}(\omega)=\frac{i}{2\pi}\int\!\text{d}u\;e^{i\eta u}\,G_{nn}(\omega+u)\,W_{nn}(u),
\end{equation}
with $G_{nn}(z)=[z-\epsilon_n-\Sigma^{\rm xc}_{nn}]^{-1}$ and $\eta\to0^{+}$. Closing the
contour in the upper half plane and collecting the $G_0$ poles that the deformation
encloses gives the CD form used in the main text,
\begin{equation}\label{eq:cdform}
\Sigma^{c,\rm CD}_{nn}(\omega)=
-\frac{1}{\pi}\sum_k w_k\frac{t_n}{t_n^{2}+u_k^{2}}W_{nn}(iu_k)
+\sum_m s_m\,W_{nm}(\zeta_m),\qquad
\zeta_m=|\epsilon_m-\omega|,\quad t_n=\omega-\epsilon_n ,
\end{equation}
where the imaginary-axis integral has been discretized on a fixed quadrature
$\{u_k,w_k\}$ (the modified Gauss--Legendre grid of the reference spectrum), and the pole
weights are
\begin{equation}\label{eq:sm}
s_m=\begin{cases}
-1, & m\ \text{occupied and}\ \epsilon_m>\omega,\\
+1, & m\ \text{unoccupied and}\ \epsilon_m<\omega,\\
0, & \text{otherwise},
\end{cases}
\end{equation}
selected relative to the Fermi energy $\epsilon_F=\tfrac12(\epsilon_{\rm HOMO}+\epsilon_{\rm LUMO})$.

\section{Static subtraction and the pole crossing}\label{sec:sub}

The two terms of Eq.~\eqref{eq:cdform} are individually discontinuous when $\omega$ approaches one of
the $\epsilon_m$: the weight $s_m$ flips and, in any grid-based evaluation, the residue
frequency jumps to a neighbouring node. The discontinuity is an artifact that stems from numerical intergation over imaginary axis and is removed by integrating the static part analytically,
\begin{equation}\label{eq:staticint}
\int_0^\infty\frac{t}{t^{2}+u^{2}}\,\text{d}u=\frac{\pi}{2}\,\mathrm{sgn}(t_0),
\end{equation}
and subtracting it from the integrand, which gives the subtracted self-energy
\begin{equation}\label{eq:sub}
\begin{aligned}
\Sigma^{c}_{n,m}(\omega)={}&-\frac{1}{\pi}\sum_k w_k\,\frac{t_m}{t_m^{2}+u_k^{2}}
\big[W_{nm}(iu_k)-W_{nm}(0)\big]
+\Big(\frac12-f_m\Big)W_{nm}(0)\\
&+s_m\,pf_m\big[W_{nm}(\zeta_m)-W_{nm}(0)\big],
\end{aligned}
\end{equation}
with $f_m=1$ ($0$) for occupied (unoccupied) $m$ and $pf_m$ the half-weight factor at the
threshold. Equation~\eqref{eq:sub} is algebraically identical to Eq.~\eqref{eq:cdform} for every
$\omega\neq\epsilon_m$, because the three terms recombine using
$-\tfrac12\mathrm{sgn}(t_m)+s_mpf_m=\tfrac12-f_m$. At $\omega=\epsilon_m$ the residue
bracket vanishes, so the sign flip of $s_m$ multiplies zero; the sum is continuous and
its fixed-$\omega$ derivative remains finite. Since $Q(z)-Q(0)=O(z^{2})$, both brackets
vanish as $z^{2}$ at the origin, so the subtracted form is also well conditioned
numerically near the threshold. All gradients reported in the main text are evaluated
from Eq.~\eqref{eq:sub}; the same subtraction is applied in the AC variant, where it is exactly
satisfied by the continued fraction (the $u\to0$ limit of the interpolant is constrained
by the lowest node).

\section{Analytic gradient of the CD self-energy}\label{sec:cdgrad}

\paragraph{Implicit differentiation and the covectors.}
The CD quasiparticle energy is the root of the same implicit equation as in the AC
variant, so
\begin{equation}\label{eq:cdgradz}
\frac{\mathrm{d}\omega_n}{\mathrm{d}\xi}
=Z_n\Big[\frac{\mathrm{d}\epsilon_n}{\mathrm{d}\xi}
+\frac{\mathrm{d}\Sigma^{x}_{nn}}{\mathrm{d}\xi}
-\frac{\mathrm{d}v^{xc}_n}{\mathrm{d}\xi}
+\Big(\frac{\mathrm{d}\Sigma^{c}_{nn}}{\mathrm{d}\xi}\Big)_{\!\omega}\Big],
\qquad Z_n=\Big[1-\frac{\partial\Sigma^{c}_{nn}}{\partial\omega}\Big]^{-1},
\end{equation}
and the whole content of the CD gradient is the fixed-$\omega$ derivative of the
correlation self-energy. Since $\Sigma^{c}_{nn}$ depends on the geometry only through
$I$, $J$ and $\{\epsilon_p\}$---the quadrature nodes, the set of active poles and the pole
weights are discrete data, held fixed under differentiation---that derivative is a linear
functional of $\delta I$, $\delta J$ and $\delta\epsilon$, i.e.\ the three correlation
covectors $\bar I^{(c)}$, $\bar J^{(c)}$, $\bar\epsilon^{(c)}$ of Eq.~(12) of the main
text. The derivation differs from the AC one only in the frequency content of the
functional, which is now spread over three channels instead of one.

\paragraph{The differentiated functional.}
Writing the static-subtracted form of Eq.~\eqref{eq:sub} as a single sum over $m$,
\begin{equation}\label{eq:cdcoef}
\Sigma^{c}_{nn}(\omega)=\sum_m\Big[\sum_k a_{k,m}\,W_{nm}(\mathrm{i}u_k)
+c^{(0)}_m\,W_{nm}(0)+a^{(\mathrm{r})}_m\,W_{nm}(\zeta_m)\Big],
\end{equation}
with
\begin{equation}\label{eq:cdcoef2}
a_{k,m}=-\frac{w_k}{\pi}\frac{t_m}{t_m^2+u_k^2},\qquad
a^{(\mathrm{r})}_m=s_m\,pf_m,\qquad
c^{(0)}_m=\tfrac12-f_m-\sum_k a_{k,m}-a^{(\mathrm{r})}_m,
\end{equation}
each channel contributes a covector of the same algebraic form. On the quadrature the
identity $-\tfrac12\mathrm{sgn}(t_m)+s_mpf_m=\tfrac12-f_m$ of Sec.~\ref{sec:sub} makes
$c^{(0)}_m$ vanish to the accuracy of the grid ($|c^{(0)}_m|\lesssim10^{-9}$); the static
channel is nevertheless retained, so that the differentiated functional is exactly the one
that is evaluated: it absorbs the quadrature residual of the identity and, at a threshold,
the mismatch between the $\tfrac12$ smoothing of $pf_m$ and the step of the quadrature
sum. The residue weight $a^{(\mathrm{r})}_m$ is frozen: $s_m$ and $pf_m$ are step functions
of $\omega$ and are not differentiated.

\paragraph{Differential of a screened matrix element.}
At a fixed frequency $z$---real, imaginary or complex---the screened interaction
$W_{nm}(z)=\sum_{PQ}I_{P,nm}[(J-Q(z))^{-1}-J^{-1}]_{PQ}I_{Q,nm}$ differentiates as
\begin{equation}\label{eq:cddw}
\begin{aligned}
\delta W_{nm}(z)&=2\sum_P\delta I_{P,nm}\big(y_{P,nm}(z)-y^0_{P,nm}\big)\\
&\quad-\sum_{P,Q}y_{P,nm}(z)\big(\delta J_{PQ}-\delta Q_{PQ}(z)\big)y_{Q,nm}(z)
+\sum_{P,Q}y^0_{P,nm}\,\delta J_{PQ}\,y^0_{Q,nm},
\end{aligned}
\end{equation}
with $y_{P,nm}(z)=\sum_Q[(J-Q(z))^{-1}]_{PQ}I_{Q,nm}$ and
$y^0_{P,nm}=\sum_Q(J^{-1})_{PQ}I_{Q,nm}$; this is the CD counterpart of Eqs.~(18)--(19) of
the main text. The polarization response is that of Eq.~(20) of the main text,
\begin{equation}\label{eq:cddq}
\delta Q_{PQ}(z)=\sum_{ia}\Big[\delta I_{P,ia}I_{Q,ia}+I_{P,ia}\delta I_{Q,ia}\Big]d_{ia}(z)
+I_{P,ia}I_{Q,ia}\,d'_{ia}(z)\,\delta\Delta_{ia},
\end{equation}
with the real-axis response evaluated with a small imaginary broadening in the retarded
denominator; at $\eta\to0$ it reduces to $d_{ia}(\omega)$ of Eq.~(4) of the main text. The
screening matrix $J-Q(z)$ is complex symmetric rather than Hermitian, so the
implementation factors it once per frequency with a complex LU decomposition and
contracts the real and imaginary parts with the transpose, never the conjugate transpose;
all covectors of Eq.~\eqref{eq:cdgradz} are real.

\paragraph{The three channels.}
Inserting Eq.~\eqref{eq:cddq} into Eq.~\eqref{eq:cddw} and reading off the coefficients of
$\delta I$, $\delta J$ and $\delta\Delta$ gives, for a channel with coefficients $c_{z,m}$
at frequency $z$, the four building blocks
\begin{equation}\label{eq:cdblocks}
\begin{aligned}
V^{(z)}_{PQ}&=\sum_m c_{z,m}\,y_{P,nm}(z)\,y_{Q,nm}(z),\\
\bar I^{(z)}_{P,nm}&=2c_{z,m}\big(y_{P,nm}(z)-y^0_{P,nm}\big),\qquad
\bar I^{(Q,z)}_{P,ia}=2\,d_{ia}(z)\sum_Q V^{(z)}_{PQ}I_{Q,ia},\\
\bar J^{(z)}_{PQ}&=\sum_m c_{z,m}\Big(y^0_{P,nm}y^0_{Q,nm}
-y_{P,nm}(z)y_{Q,nm}(z)\Big),\\
g^{(z)}_{ia}&=\sum_{P,Q}I_{P,ia}V^{(z)}_{PQ}I_{Q,ia}\,d'_{ia}(z),
\qquad \bar\epsilon^{(z)}_i=-g^{(z)}_{ia},\quad \bar\epsilon^{(z)}_a=+g^{(z)}_{ia},
\end{aligned}
\end{equation}
where the partners of the pair covector $\bar I^{(z)}_{P,nm}$ follow from $I_{P,pq}=I_{P,qp}$
(the pair covector is symmetrized, i.e.\ the two directions of the pair are averaged),
$d'_{ia}=\partial d_{ia}/\partial\Delta_{ia}$, and the signs of $\bar\epsilon^{(z)}$ follow
from $\delta\Delta_{ia}=\delta\epsilon_a-\delta\epsilon_i$. The three CD channels are then
\begin{center}
\small
\begin{tabular}{@{}llll@{}}
\toprule
channel & frequency $z$ & coefficient $c_{z,m}$ & response \\
\midrule
imaginary axis & $\mathrm{i}u_k$, every node & $a_{k,m}$ & real symmetric, reused factor \\
static & $0$ & $c^{(0)}_m$ & complex symmetric, reused factor \\
residue & $\zeta_m$, every active pole & $a^{(\mathrm{r})}_m$ & complex symmetric, one factor per pole \\
\bottomrule
\end{tabular}
\end{center}
The imaginary and static channels are ordinary sums over a fixed frequency set and are
handled exactly like the AC imaginary-axis sum of Sec.~\ref{sec:ac}, with the node
coefficient $c_{nk}$ replaced by $a_{k,m}$ and $y^{k}_{P,nm}$ replaced by
$y_{P,nm}(\mathrm{i}u_k)$. The residue channel is the only structurally new one.

\paragraph{The moving residue frequency.}
The argument of the residue term is not a fixed frequency:
\begin{equation}\label{eq:cdres}
\zeta_m=|\epsilon_m-\omega|,\qquad
\frac{\partial\zeta_m}{\partial\epsilon_m}=\mathrm{sgn}(\epsilon_m-\omega)
=-\frac{\partial\zeta_m}{\partial\omega},
\end{equation}
so at fixed $\omega$ the residue channel produces, in addition to Eq.~\eqref{eq:cdblocks}
with $c_{z,m}=a^{(\mathrm{r})}_m$, the covector
\begin{equation}\label{eq:cdres2}
\bar\epsilon^{(\zeta)}_m
=a^{(\mathrm{r})}_m\,\mathrm{sgn}(\epsilon_m-\omega)\,
\mathrm{Re}\,\frac{\mathrm{d}W_{nm}}{\mathrm{d}\zeta}\Big|_{\zeta_m},
\qquad
\frac{\mathrm{d}W_{nm}}{\mathrm{d}\zeta}\Big|_{\zeta}
=\sum_{ia}\big(I_{ia}^{T}y_{nm}(\zeta)\big)^2d'_{ia}(\zeta).
\end{equation}

\paragraph{Explicit mean-field channel.}
The kernel $t_m/(t_m^2+u_k^2)$ depends on $\epsilon_m$ through $t_m=\omega-\epsilon_m$, so
even at frozen $W$
\begin{equation}\label{eq:cdt}
\bar\epsilon^{(t)}_p=\delta_{pm}\sum_k\frac{w_k}{\pi}\,
\frac{u_k^2-t_m^2}{(t_m^2+u_k^2)^2}
\big[W_{nm}(\mathrm{i}u_k)-W_{nm}(0)\big],
\end{equation}
the CD counterpart of Eq.~(17) of the main text; omitting it produces errors of order
$10^{-2}$~a.u.\ in virtual QP gradients.

\paragraph{Assembly.}
Adding the three channels, the exact-exchange covectors of Eq.~(15) of the main text and
the explicit terms of the QP equation gives the full $GW$ covectors of Eq.~(26) of the
main text,
\begin{equation}\label{eq:cdtotal}
\bar I^{\rm GW}_{P,pq}=\bar I^{(c)}_{P,pq}+\bar I^{(x)}_{P,pq},\qquad
\bar J^{\rm GW}_{PQ}=\bar J^{(c)}_{PQ}+\bar J^{(x)}_{PQ},\qquad
\bar\epsilon^{\rm GW}_p=\delta_{pn}+\bar\epsilon^{(c)}_p,
\end{equation}
with
\begin{equation}\label{eq:cdtotal2}
\bar I^{(c)}_{P,pq}=\sum_{z}\Big[\bar I^{(z)}_{P,pq}+\bar I^{(Q,z)}_{P,pq}\Big],
\qquad
\bar\epsilon^{(c)}_p=\bar\epsilon^{(t)}_p+\bar\epsilon^{(\zeta)}_p
+\sum_{z}\bar\epsilon^{(z)}_p,
\end{equation}
and the gradient is Eq.~(27) of the main text,
\begin{equation}\label{eq:cdgradfinal}
\frac{\mathrm{d}\omega_n}{\mathrm{d}\xi}
=Z_n\Big[\sum_{P,pq}\bar I^{\rm GW}_{P,pq}\frac{\mathrm{d}I_{P,pq}}{\mathrm{d}\xi}
+\sum_{P,Q}\bar J^{\rm GW}_{PQ}\frac{\mathrm{d}J_{PQ}}{\mathrm{d}\xi}
+\sum_p\bar\epsilon^{\rm GW}_p\frac{\mathrm{d}\epsilon_p}{\mathrm{d}\xi}
-\frac{\mathrm{d}v^{xc}_n}{\mathrm{d}\xi}\Big].
\end{equation}
Two bookkeeping details of the implementation are worth recording. First, the exchange
term of the QP equation is attenuated by the fraction already contained in $\epsilon_n$:
writing $\alpha$ for the exact-exchange fraction of the reference functional, the block is
multiplied by $(1-\alpha)$, and it vanishes for $\alpha=1$ (Hartree--Fock), where the
exchange sits entirely in $\epsilon_n$. Second, when a deep level is described by the static
on-shell branch $\omega_n=\mathrm{const}_n+\mathrm{Re}\,\Sigma^{c}_{nn}(\epsilon_n)$
($Z_n=1$), the evaluation frequency moves with $\epsilon_n$ and the corresponding channel
$\bar\epsilon_n \mathrel{+}=\partial\Sigma^{c}_{nn}/\partial\omega$ is added explicitly.

\section{The CD-WAC variant}\label{sec:wac}

\paragraph{Motivation.}
CD is accurate for deep levels because it evaluates $W$ on the real axis at the physical
residue frequencies. Its gradient, however, needs $W$ at every $\zeta_m$, i.e.\ a real-axis
grid dense enough to interpolate the response. The WAC idea is to keep the real-axis
information but to represent it by a numerical fit whose evaluation cost is independent
of the number of residues.

\paragraph{Construction.}
Following Panad\'es-Barrueta and Golze [J. Chem. Theory Comput. \textbf{19}, 5450
(2023)], the residue sum of Eq.~\eqref{eq:cdform} is replaced by a fit of the screened
interaction that is built from reference values at
\begin{itemize}
\item the imaginary-axis quadrature nodes (all 20 by default, or a thinned subset of
$n_{\rm imag}$ nodes when \texttt{wac\_n\_imag} is set), and
\item a small set of $n_{\rm real}$ near-real reference frequencies
$r_p+\mathrm{i}\delta$ with $0<r_1<\cdots<r_{n_{\rm real}}\le\omega_{\max}$ (default
$n_{\rm real}=20$, $\omega_{\max}=50$~Ha, $\delta=10^{-3}$~Ha),
\end{itemize}
and it is a Pad\'e approximant written as a Thiele continued fraction. Because the RPA
response of Eq.~(4) of the main text is even in the frequency, the interpolation variable
is $u=z^{2}$ and the reference values are $W_{nm}(z_p)$ at $u_p=z_p^{2}$:
\begin{equation}\label{eq:wacfit}
\mathcal{W}_{nm}(u)=A_0+\frac{u-u_0}{A_1+\dfrac{u-u_1}{A_2+\cdots+\dfrac{u-u_{N-2}}{A_{N-1}}}},
\qquad u=z^{2},
\end{equation}
whose $N$ coefficients ($N=n_{\rm ref}$ in the production setup) are the Thiele
reciprocal differences of the reference values (Sec.~\ref{sec:wacgrad}); here
$n_{\rm ref}=n_{\rm imag}+n_{\rm real}$ is the total number of reference frequencies---40
with the defaults---and not an orbital count. Because $W$ is a dense matrix, one such continued fraction
is built for every matrix element $W_{nm}$ of the target row, all on the same node set and
with the reference solves shared between them (Sec.~\ref{sec:wacgrad}). The continued fraction passes through every reference value, so the imaginary-axis data are
reproduced almost exactly while the real-axis fitting points anchor the fit inside the window that
brackets the residue frequencies of the states of interest. The cost is
fixed by the number of reference frequencies,
so evaluating $\mathcal{W}_{nm}$ at any frequency costs $O(n_{\rm ref})$ instead of
$O(N_{\rm occ}N_{\rm vir})$ residue rows. The
CD-WAC gradient costs the same as the AC gradient within a few percent, while the CD gradient costs an order of magnitude more for systems with more than
$\sim$30 occupied orbitals since there are more residues to evaluate.

\section{Analytic gradient of the CD-WAC self-energy}\label{sec:wacgrad}

\paragraph{The differentiated functional.}
The CD-WAC energy functional evaluated by the production code is
\begin{equation}\label{eq:wacfun}
\Sigma^{c,\rm WAC}_{nn}(\omega)
=-\frac{1}{\pi}\sum_{m,k}w_k\,\frac{t_m}{t_m^2+u_k^2}\,W_{nm}(\mathrm{i}u_k)
+\sum_m w_m\,\mathrm{Re}\,\mathcal{P}_m(\zeta_m),
\qquad w_m=\varsigma_m\,pf_m,
\end{equation}
with $\varsigma_m=+1$ ($-1$) for a virtual (occupied) pole selected relative to the Fermi
level, exactly as in Sec.~\ref{sec:cd}. The imaginary-axis term is the same as in CD; the residue
sum of Eq.~\eqref{eq:cdform} is replaced by the fit $\mathcal{P}_m$, and no explicit
$W(0)$ subtraction appears: the residue channel is now the analytic function
$\mathcal{P}_m$ evaluated at the exact frequency, and the pole weight is frozen as in
Sec.~\ref{sec:sub}. Away from a pole threshold the raw and subtracted decompositions
differ only by the $O(10^{-9})$ residual discussed in Sec.~\ref{sec:cdgrad}, i.e.\ by less
than the accuracy of the fit. The gradient is again Eq.~\eqref{eq:cdgradz}, with $Z_n$
containing the derivative of the fit; its whole content is the fixed-$\omega$ derivative of
Eq.~\eqref{eq:wacfun}.

\paragraph{Reference set and Pad\'e variable.}
Following Sec.~\ref{sec:wac}, the fit is built on
\begin{equation}\label{eq:wacref}
z_p=\begin{cases}
\mathrm{i}u_k, & p=1,\dots,n_{\rm imag},\\[2pt]
r_p+\mathrm{i}\delta, & p=n_{\rm imag}+1,\dots,n_{\rm ref},
\end{cases}
\qquad u_p=z_p^2,
\end{equation}
where the first group are the (possibly thinned) imaginary-axis quadrature nodes and
$0<r_1<\cdots<r_{n_{\rm real}}\le\omega_{\max}$ the near-real reference frequencies of
Sec.~\ref{sec:wac}; the reference set therefore contains
$n_{\rm ref}=n_{\rm imag}+n_{\rm real}$ frequencies (40 with the defaults), which is the
order of the continued fraction and the size of the Pad\'e Jacobian. The interpolation
variable is $u=z^2$ (Sec.~\ref{sec:wac}), so the
imaginary and near-real references live in the same continued fraction and the residue is
evaluated at $u=(\zeta_m+\mathrm{i}\eta)^2$, with $\eta$ the response broadening
(\texttt{cdgw\_eta}). The reference values
\begin{equation}\label{eq:wacvals}
W_{nm}(z_p)=\sum_{P,Q}I_{P,nm}\big[(J-Q(z_p))^{-1}-J^{-1}\big]_{PQ}I_{Q,nm}
\end{equation}
are obtained from the geometry-fixed complex factors of $I-\chi_0(z_p)$ in the
metric-transformed representation, so they cost one triangular solve per target orbital
and reference frequency.

\paragraph{Element-wise fit of the dense matrix.}
The screened interaction is a dense matrix, and the continuation is performed
\emph{element by element}: for a target orbital $n$ one continued fraction
$\mathcal{P}_m$ is built for every column $m$ from the samples $W_{nm}(z_p)$ of that
element, all on the same shared node set $\{u_p\}$. The Jacobian rows $\alpha_{m,p}$,
$\nu_{m,p}$ of Eq.~\eqref{eq:wacalpha} are therefore per-element quantities---the column
index in $\mathcal{P}_m$ carries precisely this---and the greedy ordering of the reference
points is selected separately for every element. What is shared is the expensive part: the
reference factors $(I-\chi_0(z_p))^{-1}$ and the node set are built once per geometry and
reused by every element of every target orbital, so a target orbital costs $n_{\rm ref}$
triangular solves (each with $n_{\rm mo}$ right-hand sides) plus one scalar interpolation
per active pole, of $O(n_{\rm ref}^3)$ work and $n_{\rm ref}$ stored coefficients per
element, rather than $n_{\rm ref}$ factorizations per element. Since $W$ is symmetric only
the target row is needed, so the number of fits is at most $n_{\rm mo}$ per target orbital
rather than $n_{\rm mo}^2$, and the same count applies to the CD-WAC entry list
$\{(\alpha_{m,p},\beta_m)\}$ consumed by the pullback.

\paragraph{Chain rule: value and node channels.}
The geometry enters Eq.~\eqref{eq:wacfun} both through the reference values and through
the arguments of the fit (the evaluation node $\zeta_m$ and, when $\omega_{\max}$ is not
pinned, the reference nodes $u_p$). Hence
\begin{equation}\label{eq:wacchain}
\Big(\frac{\partial\Sigma^{c,\rm WAC}_{nn}}{\partial\xi}\Big)_{\!\omega}
=\sum_m w_m\,\mathrm{Re}\Big[
\sum_p\alpha_{m,p}\frac{\mathrm{d}W_{nm}(z_p)}{\mathrm{d}\xi}
+\frac{\partial\mathcal{P}_m}{\partial\zeta}\frac{\mathrm{d}\zeta_m}{\mathrm{d}\xi}
+\sum_p\nu_{m,p}\frac{\mathrm{d}u_p}{\mathrm{d}\xi}\Big],
\end{equation}
with the Pad\'e Jacobian rows
\begin{equation}\label{eq:wacalpha}
\alpha_{m,p}=\frac{\partial\mathcal{P}_m}{\partial W_{nm}(z_p)},\qquad
\nu_{m,p}=\frac{\partial\mathcal{P}_m}{\partial u_p}.
\end{equation}
The three terms are the reference-value channel, the evaluation-node channel and the
reference-node channel. The second is the direct CD-WAC analogue of Eq.~\eqref{eq:cdres2};
the third vanishes for a pinned $\omega_{\max}$ and is the analogue of the grid-motion
channel of Sec.~\ref{sec:cdgrad}.

\paragraph{Pad\'e Jacobian.}
Writing the continued fraction as
\begin{equation}\label{eq:waceval}
\mathcal{P}(u)=A_0+\frac{u-u_0}{A_1+\dfrac{u-u_1}{A_2+\cdots+\dfrac{u-u_{N-2}}{A_{N-1}}}},
\end{equation}
its coefficients are the Thiele reciprocal differences
\begin{equation}\label{eq:wacjeff}
\begin{aligned}
&A_0=W_0,\qquad \rho_i^{(0)}=W_i,\qquad
\rho_i^{(j)}=\frac{u_i-u_{j-1}}{\rho_i^{(j-1)}}-A_j\quad(1\le j<i),\\
&A_i=\frac{u_i-u_{i-1}}{\rho_i^{(i-1)}}\quad(1\le i\le N-1),
\end{aligned}
\end{equation}
so every $A_i$ is a smooth function of $\{W_0,\dots,W_i,u_0,\dots,u_i\}$. Differentiating
Eq.~\eqref{eq:wacjeff} with the same forward-mode algebra used for the AC Jacobian in
Sec.~\ref{sec:ac}---carrying the
value, the first and second derivatives in $u$ and the derivatives with respect to every
parameter---gives $\partial A_i/\partial p$ for $p\in\{W_p,u_p\}$, and differentiating
Eq.~\eqref{eq:waceval} gives $\partial\mathcal{P}/\partial A_i$ together with the explicit
$\partial\mathcal{P}/\partial u_p$ of the numerators $u-u_p$. Therefore
\begin{equation}\label{eq:wacjac}
\frac{\partial\mathcal{P}}{\partial W_p}
=\sum_i\frac{\partial\mathcal{P}}{\partial A_i}\frac{\partial A_i}{\partial W_p},
\qquad
\frac{\partial\mathcal{P}}{\partial u_p}
=\sum_i\frac{\partial\mathcal{P}}{\partial A_i}\frac{\partial A_i}{\partial u_p}
+\frac{\partial\mathcal{P}}{\partial u_p}\Big|_{\rm num},
\end{equation}
which is Eq.~\eqref{eq:jac} generalized from sample parameters to the pair
(value, node). The Jacobian is dense in the reference index; $\alpha_{m,p}$ and
$\nu_{m,p}$ are $n_{\rm ref}$-dimensional rows, built once per geometry, target orbital
and active pole. The greedy ordering of the reference points (largest $|W|$ first, then
largest deviation from the current approximant) is a discrete choice and is held fixed
under differentiation; the derivatives are mapped back to the original reference order.

\paragraph{Reference-value channel.}
A reference value contributes through $\alpha_{m,p}$ exactly like a fixed-frequency CD
residue, so the building blocks of Eq.~\eqref{eq:cdblocks} apply with $z=z_p$ and
$c_{z,m}=w_m\alpha_{m,p}$, with two simplifications: only the real part of the complex
response survives, because the functional is $\mathrm{Re}\,\mathcal{P}$, and the reference
values are linear in $(J-Q)^{-1}$, so the same complex-symmetric solve provides
$y_{nm}(z_p)$ and $y^0_{nm}$:
\begin{equation}\label{eq:wacpull}
\begin{aligned}
\bar I^{(p)}_{P,nm}&=2w_m\alpha_{m,p}\big(\mathrm{Re}\,y_{P,nm}(z_p)-y^0_{P,nm}\big),\\
\bar J^{(p)}_{PQ}&=\sum_m w_m\alpha_{m,p}\Big(y^0_{P,nm}y^0_{Q,nm}
-\mathrm{Re}\,y_{P,nm}(z_p)y_{Q,nm}(z_p)\Big),\\
g^{(p)}_{ia}&=\sum_{P,Q}I_{P,ia}V^{(p)}_{PQ}I_{Q,ia}\,\mathrm{Re}\,d'_{ia}(z_p),
\qquad \bar\epsilon^{(p)}_i=-g^{(p)}_{ia},\quad\bar\epsilon^{(p)}_a=+g^{(p)}_{ia},
\end{aligned}
\end{equation}
with $V^{(p)}_{PQ}=\sum_m w_m\alpha_{m,p}\,\mathrm{Re}
[y_{P,nm}(z_p)y_{Q,nm}(z_p)]$. These are the CD-WAC counterparts of the residue channel of
Eq.~\eqref{eq:cdblocks}: the exact residue row $y_{nm}(\zeta_m)$ is replaced by the
reference rows $y_{nm}(z_p)$, weighted by the Pad\'e coefficients.

\paragraph{Evaluation-node channel.}
The evaluation node moves with the spectrum,
\begin{equation}\label{eq:wacdu}
\frac{\mathrm{d}u}{\mathrm{d}\xi}=\frac{\partial u}{\partial\zeta}\frac{\mathrm{d}\zeta_m}{\mathrm{d}\xi},
\qquad\frac{\partial u}{\partial\zeta}=2(\zeta_m+\mathrm{i}\eta),
\qquad\frac{\mathrm{d}\zeta_m}{\mathrm{d}\epsilon_m}=\mathrm{sgn}(\epsilon_m-\omega),
\end{equation}
so the explicit channel is
\begin{equation}\label{eq:wacbeta}
\bar\epsilon^{(\zeta)}_m
=w_m\,\mathrm{sgn}(\epsilon_m-\omega)\,
\mathrm{Re}\Big[2(\zeta_m+\mathrm{i}\eta)\frac{\partial\mathcal{P}_m}{\partial u}\Big],
\end{equation}
the CD-WAC counterpart of Eq.~\eqref{eq:cdres2}: the exact $\mathrm{d}W/\mathrm{d}\zeta$ is
replaced by the derivative of the fit. The same derivative with
$\partial\zeta_m/\partial\omega=-\mathrm{sgn}(\epsilon_m-\omega)$ gives
$\partial\Sigma^{c}_{nn}/\partial\omega$ and hence $Z_n$.

\paragraph{Reference-node channel.}
When $\omega_{\max}$ is not pinned it is set by the spectrum,
$\omega_{\max}=\epsilon_{\rm last}-\epsilon_{\rm first}$ (when the span exceeds 1~Ha), and
the reference frequencies move:
\begin{equation}\label{eq:wacnode}
\frac{\mathrm{d}u_p}{\mathrm{d}\xi}
=\frac{\partial u_p}{\partial\omega_{\max}}
\big(\dot\epsilon_{\rm last}-\dot\epsilon_{\rm first}\big),
\qquad
\frac{\mathrm{d}W_{nm}(z_p)}{\mathrm{d}\xi}\Big|_{\rm node}
=\frac{\partial W_{nm}}{\partial z_p}\frac{\partial z_p}{\partial\omega_{\max}}
\big(\dot\epsilon_{\rm last}-\dot\epsilon_{\rm first}\big),
\end{equation}
with $\partial u_p/\partial\omega_{\max}=2z_p\,\partial z_p/\partial\omega_{\max}$ and
$\partial W_{nm}/\partial z_p=\sum_{ia}(I_{ia}^{T}y_{nm}(z_p))^2d'_{ia}(z_p)$. Collecting
both into a single covector on the two orbitals that define $\omega_{\max}$,
\begin{equation}\label{eq:wacnodecoef}
\bar\epsilon^{(\omega_{\max})}_{\rm last}=\sum_m w_m\sum_p\Big[
\nu_{m,p}\frac{\partial u_p}{\partial\omega_{\max}}
+\alpha_{m,p}\,\mathrm{Re}\frac{\partial W_{nm}}{\partial z_p}
\frac{\partial z_p}{\partial\omega_{\max}}\Big],
\qquad
\bar\epsilon^{(\omega_{\max})}_{\rm first}=-\bar\epsilon^{(\omega_{\max})}_{\rm last}.
\end{equation}
In the production setup $\omega_{\max}$ is pinned (\texttt{wac\_omega\_max} $=50$~Ha), so
all these derivatives vanish and only Eq.~\eqref{eq:wacbeta} remains; the channel is
retained for the automatic mode, where it is the CD-WAC analogue of the grid-motion
channel of Sec.~\ref{sec:cdgrad}.

\paragraph{Assembly and cost.}
Adding the CD imaginary channel of Eq.~\eqref{eq:cdblocks} with $c_{z,m}=a_{k,m}$, the
exchange and static terms of the QP equation, and the two CD-WAC channels
Eqs.~\eqref{eq:wacpull}--\eqref{eq:wacnodecoef}, the total covectors are those of
Eqs.~\eqref{eq:cdtotal}--\eqref{eq:cdtotal2} and the gradient is
Eq.~\eqref{eq:cdgradfinal}. The algorithmic point is that the reference factors
$(I-\chi_0(z_p))^{-1}$ are shared by every target orbital and every Newton iteration, so
the residue channel of the CD-WAC gradient costs $n_{\rm ref}$ complex solves per target
orbital and become nearly independent from the number of residues(unlike in CD where the cost of evaluating each orbital increases significantly as the number of residues increases), at
the price of the fit's controlled error.

\section{Final relaxed geometries of the ground and excited states}\label{sec:geometries}

\subsection{DA1, ground state (PBEh($\alpha=0.45$))}

Table~\ref{tab:da1_s0_geom_si} lists the PBEh($\alpha=0.45$)/def2-TZVP ground-state
minimum of DA1 from which the excited-state relaxation starts. The atom labels follow
Fig.~1 of the main text (C1--C14, H15--H27, N28--N30).

\begin{table}[H]
\centering
\tabfont
\renewcommand{\arraystretch}{0.95}
\caption{Final PBEh($\alpha=0.45$)/def2-TZVP ground-state geometry of DA1. Coordinates
in \AA. The structure is the 14-step converged minimum with
$E_{\rm total}=-704.9938987000$~Ha and an rms gradient of
\num{6.2e-5}~Ha/Bohr (maximum \num{1.7e-4}~Ha/Bohr). It is the ground-state structure
from which the S$_1$ relaxation starts.}
\label{tab:da1_s0_geom_si}
\begin{tabular}{lrrr}
\toprule
Atom & $x$ & $y$ & $z$ \\
\midrule
C1 &     8.599119 &     4.015807 &     5.385491 \\
C2 &     9.348350 &     4.362381 &     4.093247 \\
C3 &    10.023779 &     5.673891 &     4.235608 \\
C4 &     9.161003 &     6.765253 &     4.748037 \\
C5 &     8.422563 &     6.282869 &     6.002088 \\
C6 &     3.826193 &     5.177394 &     4.156355 \\
C7 &     4.438732 &     3.978055 &     4.459207 \\
C8 &     5.711837 &     3.942781 &     4.979327 \\
C9 &     6.428198 &     5.116172 &     5.210491 \\
C10 &     5.801135 &     6.319608 &     4.900949 \\
C11 &     4.519860 &     6.343224 &     4.388918 \\
C12 &    11.331347 &     5.847130 &     3.973724 \\
C13 &    12.151314 &     4.780080 &     3.505589 \\
C14 &    11.968249 &     7.110544 &     4.140797 \\
H15 &     9.317566 &     3.858590 &     6.188789 \\
H16 &     8.072689 &     3.081425 &     5.248912 \\
H17 &     8.612703 &     4.438405 &     3.290258 \\
H18 &    10.049487 &     3.575934 &     3.830599 \\
H19 &     8.422145 &     7.003287 &     3.981070 \\
H20 &     9.731822 &     7.665525 &     4.954841 \\
H21 &     7.761185 &     7.057011 &     6.371985 \\
H22 &     9.152144 &     6.098248 &     6.789775 \\
H23 &     2.827447 &     5.200330 &     3.752337 \\
H24 &     3.914405 &     3.048397 &     4.300715 \\
H25 &     6.134736 &     2.986837 &     5.237747 \\
H26 &     6.304201 &     7.259032 &     5.052752 \\
H27 &     4.067457 &     7.296491 &     4.163016 \\
N28 &     7.700055 &     5.066153 &     5.777212 \\
N29 &    12.803367 &     3.919884 &     3.132756 \\
N30 &    12.471698 &     8.126277 &     4.278408 \\
\bottomrule
\end{tabular}
\end{table}

\clearpage
\subsection{DA1, PBEh($\alpha=0.45$)/TDA}

Table~\ref{tab:da1_geom_si} lists the converged PBEh($\alpha=0.45$)/TDA
AC-GW-BSE/def2-TZVP geometry of DA1 discussed in Sec.~4.2 of the main text.
The atom labels follow Fig.~1 of the main text (C1--C14, H15--H27, N28--N30).

\begin{table}[H]
\centering
\tabfont
\renewcommand{\arraystretch}{0.95}
\caption{Final PBEh($\alpha=0.45$)/TDA AC-GW-BSE geometry of DA1 (def2-TZVP).
Coordinates in \AA. The structure is the 17-step converged minimum with
$E_{\rm total}=-704.8641324846$~Ha, $\Omega=0.1050132262$~Ha ($2.8576$~eV),
and an RMS gradient of \num{7.8e-5}~Ha/Bohr.}
\label{tab:da1_geom_si}
\begin{tabular}{lrrr}
\toprule
Atom & $x$ & $y$ & $z$ \\
\midrule
  C1 &     8.359284 &     3.991948 &     5.359327 \\
  C2 &     9.352117 &     4.317698 &     4.246891 \\
  C3 &    10.042148 &     5.600183 &     4.496155 \\
  C4 &     9.170220 &     6.733535 &     4.871505 \\
  C5 &     8.188128 &     6.308892 &     5.952770 \\
  C6 &     3.582265 &     5.213561 &     4.154420 \\
  C7 &     4.206106 &     4.000284 &     4.394156 \\
  C8 &     5.498400 &     3.958657 &     4.835862 \\
  C9 &     6.211381 &     5.153448 &     5.068704 \\
 C10 &     5.555920 &     6.379934 &     4.829042 \\
 C11 &     4.268448 &     6.396967 &     4.372117 \\
 C12 &    11.279632 &     5.862125 &     3.841159 \\
 C13 &    11.998400 &     4.819905 &     3.244547 \\
 C14 &    11.814784 &     7.155253 &     3.835325 \\
 H15 &     8.888962 &     3.856848 &     6.298233 \\
 H16 &     7.798634 &     3.099141 &     5.126039 \\
 H17 &     8.789975 &     4.311134 &     3.295581 \\
 H18 &    10.056880 &     3.491269 &     4.182709 \\
 H19 &     8.587520 &     7.125484 &     4.018574 \\
 H20 &     9.748727 &     7.573140 &     5.250439 \\
 H21 &     7.487294 &     7.085111 &     6.227618 \\
 H22 &     8.745045 &     6.017536 &     6.840055 \\
 H23 &     2.565452 &     5.236689 &     3.797443 \\
 H24 &     3.671167 &     3.077824 &     4.237527 \\
 H25 &     5.949904 &     3.005642 &     5.040347 \\
 H26 &     6.083450 &     7.308334 &     4.948845 \\
 H27 &     3.792497 &     7.342430 &     4.169142 \\
 N28 &     7.478188 &     5.127378 &     5.516372 \\
 N29 &    12.535251 &     3.917157 &     2.772896 \\
 N30 &    12.194378 &     8.241788 &     3.867597 \\
\bottomrule
\end{tabular}
\end{table}

\clearpage
\subsection{DMABN, ground state (PBE0)}

Table~\ref{tab:dmabn_s0_geom_si} lists the PBE0/def2-SVP ground-state minimum of DMABN
from which the two excited-state relaxations start; it is the planar reference used for
the twist angles of Tables~\ref{tab:dmabn_pbe0_geom_si} and~\ref{tab:dmabn_pbeh_geom_si}.
The atom labelling is that of Table~\ref{tab:dmabn_pbe0_geom_si}.

\begin{table}[H]
\centering
\tabfont
\renewcommand{\arraystretch}{0.95}
\caption{Final PBE0/def2-SVP ground-state geometry of DMABN. Coordinates in \AA. The
structure is the 50-step converged minimum with $E_{\rm total}=-457.5909679788$~Ha and
an rms gradient of \num{4.1e-5}~Ha/Bohr; the dimethylamino group is planar with respect
to the ring (twist \SI{0.5}{\degree}).}
\label{tab:dmabn_s0_geom_si}
\begin{tabular}{lrrr}
\toprule
Atom & $x$ & $y$ & $z$ \\
\midrule
  C1   &       1.526597 &      -0.011608 &       0.028048 \\
  C2   &       0.775285 &       1.173599 &       0.073967 \\
  C3   &      -0.608492 &       1.142801 &       0.068435 \\
  C4   &      -1.313596 &      -0.084612 &       0.016669 \\
  C5   &      -0.546070 &      -1.274084 &      -0.029415 \\
  C6   &       0.837535 &      -1.233809 &      -0.023850 \\
  C7   &       2.954227 &       0.025246 &       0.034011 \\
  N8   &       4.114849 &       0.055457 &       0.038878 \\
  N9   &      -2.681932 &      -0.119649 &       0.011934 \\
  C10  &      -3.435586 &       1.108056 &       0.047188 \\
  H11  &      -3.239723 &       1.691764 &       0.964644 \\
  H12  &      -3.213078 &       1.757271 &      -0.818518 \\
  H13  &      -4.507125 &       0.876709 &       0.022233 \\
  C14  &      -3.372189 &      -1.383714 &      -0.034373 \\
  H15  &      -4.454352 &      -1.207020 &      -0.019371 \\
  H16  &      -3.138356 &      -1.954042 &      -0.951404 \\
  H17  &      -3.125577 &      -2.023820 &       0.831527 \\
  H18  &       1.291641 &       2.134875 &       0.115134 \\
  H19  &      -1.149323 &       2.087595 &       0.106220 \\
  H20  &      -1.037481 &      -2.245350 &      -0.071189 \\
  H21  &       1.402840 &      -2.167312 &      -0.060741 \\
\bottomrule
\end{tabular}
\end{table}

\clearpage
\subsection{DMABN, PBE0/TDA}

Table~\ref{tab:dmabn_pbe0_geom_si} lists the converged PBE0/TDA AC-GW-BSE/def2-SVP
geometry of the DMABN charge-transfer minimum discussed in Sec.~4.2 of the main text
(twist \SI{84.4}{\degree}). The atom labels are C1--C6 for the ring (C1 carries the
cyano group, C4 the dimethylamino group), C7 and N8 for the cyano group, N9 for the
amine nitrogen, C10 and C14 for the two methyl carbons, and H11--H13, H15--H17 and
H18--H21 for their hydrogens.

\begin{table}[H]
\centering
\tabfont
\renewcommand{\arraystretch}{0.95}
\caption{Final PBE0/TDA AC-GW-BSE geometry of DMABN (def2-SVP, AC-$G_0W_0$ with 24 imaginary-axis samples, $27\times52$ pinned active space). Coordinates in \AA. The structure is the 54-step converged minimum with $E_{\rm total}=-457.4430606400$~Ha, $\Omega=0.089650795$~Ha ($2.4395$~eV), and an rms gradient of \num{5.3e-5}~Ha/Bohr.}
\label{tab:dmabn_pbe0_geom_si}
\begin{tabular}{lrrr}
\toprule
Atom & $x$ & $y$ & $z$ \\
\midrule
  C1   &       1.309215 &       0.165477 &      -0.051709 \\
  C2   &       0.459874 &       1.168420 &       0.508759 \\
  C3   &      -0.862594 &       0.922673 &       0.788535 \\
  C4   &      -1.426144 &      -0.397471 &       0.565345 \\
  C5   &      -0.589325 &      -1.382507 &      -0.085315 \\
  C6   &       0.729341 &      -1.117528 &      -0.340425 \\
  C7   &       2.661921 &       0.434987 &      -0.343392 \\
  N8   &       3.781585 &       0.656222 &      -0.583352 \\
  N9   &      -2.776598 &      -0.335243 &       0.124024 \\
  C10  &      -3.122909 &      -0.015876 &      -1.252891 \\
  H11  &      -2.197924 &       0.114713 &      -1.824251 \\
  H12  &      -3.721871 &      -0.842280 &      -1.662295 \\
  H13  &      -3.724821 &       0.904552 &      -1.265225 \\
  C14  &      -3.841853 &      -0.530403 &       1.079614 \\
  H15  &      -4.831347 &      -0.440405 &       0.613077 \\
  H16  &      -3.699338 &      -1.521392 &       1.531409 \\
  H17  &      -3.715059 &       0.217354 &       1.875122 \\
  H18  &       0.883476 &       2.147085 &       0.748138 \\
  H19  &      -1.484912 &       1.710261 &       1.224553 \\
  H20  &      -1.003880 &      -2.366323 &      -0.326655 \\
  H21  &       1.364599 &      -1.899825 &      -0.762554 \\
\bottomrule
\end{tabular}
\end{table}

\clearpage
\subsection{DMABN, S$_2$ (PBE0/TDA)}

Table~\ref{tab:dmabn_s2_geom_si} lists the second singlet minimum of DMABN obtained with
the same PBE0/TDA setup, followed by index so that the HOMO$\rightarrow$LUMO$+1$ state
could be relaxed independently of the charge-transfer state above. The structure is
almost planar (twist \SI{7.2}{\degree}) with a completely equalized ring and is the
second reference angle marked in Fig.~7 of the main text.

\begin{table}[h]
\centering
\tabfont
\renewcommand{\arraystretch}{0.95}
\caption{Final PBE0/TDA AC-GW-BSE geometry of the second singlet minimum of DMABN
(def2-SVP, AC-$G_0W_0$ with 24 imaginary-axis samples, $27\times52$ pinned active
space). Coordinates in \AA. The structure is the 33-step converged minimum with
$E_{\rm total}=-457.4369609878$~Ha, $\Omega=0.147167548$~Ha ($4.0046$~eV, dominated by
the HOMO$\rightarrow$LUMO$+1$ transition) and an rms gradient of
\num{3.1e-4}~Ha/Bohr; the atom labelling is that of Table~\ref{tab:dmabn_pbe0_geom_si}.}
\label{tab:dmabn_s2_geom_si}
\begin{tabular}{lrrr}
\toprule
Atom & $x$ & $y$ & $z$ \\
\midrule
  C1   &       1.539039 &       0.000534 &       0.025518 \\
  C2   &       0.814908 &       1.116613 &      -0.474116 \\
  C3   &      -0.610594 &       1.071285 &      -0.488305 \\
  C4   &      -1.270205 &      -0.090659 &      -0.000002 \\
  C5   &      -0.545000 &      -1.206921 &       0.502059 \\
  C6   &       0.880244 &      -1.159776 &       0.514211 \\
  C7   &       2.958667 &       0.045764 &       0.036252 \\
  N8   &       4.121717 &       0.082943 &       0.045190 \\
  N9   &      -2.653645 &      -0.136250 &      -0.013196 \\
  C10  &      -3.425118 &       0.908789 &      -0.654360 \\
  H11  &      -3.369769 &       1.847210 &      -0.079082 \\
  H12  &      -3.043409 &       1.099033 &      -1.666737 \\
  H13  &      -4.473261 &       0.598294 &      -0.720680 \\
  C14  &      -3.365729 &      -1.227682 &       0.619143 \\
  H15  &      -4.435739 &      -0.995900 &       0.646606 \\
  H16  &      -3.220818 &      -2.166037 &       0.059880 \\
  H17  &      -3.004999 &      -1.376723 &       1.646274 \\
  H18  &       1.350483 &       1.996709 &      -0.828075 \\
  H19  &      -1.168237 &       1.952596 &      -0.799895 \\
  H20  &      -1.049853 &      -2.121902 &       0.806406 \\
  H21  &       1.465560 &      -2.002669 &       0.879583 \\
\bottomrule
\end{tabular}
\end{table}

\subsection{DMABN, PBEh($\alpha=0.45$)/TDA}

Table~\ref{tab:dmabn_pbeh_geom_si} lists the converged PBEh($\alpha=0.45$)/TDA
AC-GW-BSE/def2-SVP geometry of the same state (twist \SI{90.0}{\degree}), with the
atom labelling of Table~\ref{tab:dmabn_pbe0_geom_si}.

\begin{table}[H]
\centering
\tabfont
\renewcommand{\arraystretch}{0.95}
\caption{Final PBEh($\alpha=0.45$)/TDA AC-GW-BSE geometry of DMABN (def2-SVP, AC-$G_0W_0$ with 24 imaginary-axis samples, $27\times52$ pinned active space). Coordinates in \AA. The structure is the 11-step converged minimum with $E_{\rm total}=-457.4682003383$~Ha, $\Omega=0.131444296$~Ha ($3.5768$~eV), and an rms gradient of \num{8.2e-5}~Ha/Bohr.}
\label{tab:dmabn_pbeh_geom_si}
\begin{tabular}{lrrr}
\toprule
Atom & $x$ & $y$ & $z$ \\
\midrule
  C1   &       1.508703 &      -0.011934 &       0.030665 \\
  C2   &       0.739202 &       1.146394 &       0.363313 \\
  C3   &      -0.622169 &       1.124484 &       0.361269 \\
  C4   &      -1.307212 &      -0.087039 &       0.017964 \\
  C5   &      -0.556186 &      -1.260674 &      -0.318605 \\
  C6   &       0.804348 &      -1.209322 &      -0.308239 \\
  C7   &       2.913395 &       0.025244 &       0.036539 \\
  N8   &       4.069765 &       0.055814 &       0.041296 \\
  N9   &      -2.694201 &      -0.123328 &       0.011133 \\
  C10  &      -3.413193 &       0.202291 &      -1.198295 \\
  H11  &      -3.146880 &       1.226958 &      -1.471004 \\
  H12  &      -3.046504 &      -0.459105 &      -1.987075 \\
  H13  &      -4.492552 &       0.101499 &      -1.071627 \\
  C14  &      -3.406298 &      -0.486541 &       1.213902 \\
  H15  &      -4.488176 &      -0.440477 &       1.077538 \\
  H16  &      -3.090962 &      -1.496779 &       1.488157 \\
  H17  &      -3.080813 &       0.191553 &       2.006537 \\
  H18  &       1.256337 &       2.065957 &       0.624163 \\
  H19  &      -1.189095 &       2.016701 &       0.617066 \\
  H20  &      -1.072426 &      -2.181733 &      -0.579504 \\
  H21  &       1.372081 &      -2.099937 &      -0.564230 \\
\bottomrule
\end{tabular}
\end{table}